\documentclass[lettersize,journal]{IEEEtran}
\usepackage{amsmath,amsfonts}
\usepackage{algorithmic}
\usepackage{algorithm}
\usepackage{makecell}
\usepackage{array}
\usepackage{textcomp}
\usepackage{stfloats}
\usepackage{url}
\usepackage{hyperref}
\usepackage{verbatim}
\usepackage{graphicx}
\usepackage{cite}
\usepackage{xcolor}
\usepackage{multicol}
\usepackage{multirow}
\usepackage{tikz}
\usepackage{pgfplots}
\usepackage{pgfplotstable}
\pgfplotsset{compat=1.7}
\usetikzlibrary{patterns}
\usepackage{subcaption}
\usepackage[]{mdframed}
\usepackage{ragged2e}

\begin{document}


\title{CodeT5-RNN: Reinforcing Contextual Embeddings for Enhanced Code Comprehension}

\author{Md Mostafizer Rahman, Ariful Islam Shiplu, Yutaka Watanobe, Md Faizul Ibne Amin, Syed Rameez Naqvi, and Fang Liu

\thanks{Md Mostafizer Rahman is with the Lucy Family Institute for Data \& Society, University of Notre Dame, IN, USA (e-mail: mostafiz26@gmail.com, mrahman3@nd.edu).}

\thanks{Ariful Islam Shiplu is with the Department of Computer Science and Engineering, Dhaka University of Engineering \& Technology, Gazipur, Bangladesh (e-mail: shipluarifulislam@gmail.com).}

\thanks{Yutaka Watanobe is with the Department of Computer and Information Systems, The University of Aizu, Aizu-Wakamatsu, Fukushima 965-8580, Japan (e-mail: yutaka@u-aizu.ac.jp).}

\thanks{Md Faizul Ibne Amin with the  Department of Computer and Information Systems, The University of Aizu, Aizu-Wakamatsu, Fukushima 965-8580, Japan (e-mail: aminfaizul007@gmail.com).}

\thanks{Syed Rameez Naqvi, is with the  Department of Computer Science, Tulane University, New Orleans,  LA, USA (e-mail: snaqvi@tulane.edu).}

\thanks{ Fang Liu is with the Department of Applied and Computational Mathematics and Statistics, and Lucy Family Institute for Data \& Society, University of Notre Dame, IN, USA (e-mail: fang.liu.131@nd.edu).}
}



\maketitle

\begin{abstract}

Large Language Models (LLMs) have achieved remarkable success in multifarious coding-related affairs, particularly in code understanding tasks such as code classification, clone detection, language identification, and algorithm recognition. 
Given their practical impact on education, software engineering, and intelligent development tools, LLM-based code analysis has attracted substantial attention from both academia and industry. Despite these advances, contextual embeddings generated by LLMs exhibit strong positional inductive biases, which can limit their ability to fully capture long-range, order-sensitive dependencies in highly structured source code. Consequently, how to further refine and enhance LLM embeddings for improved code understanding remains an open research question.
To address this gap, we propose a hybrid LLM–RNN framework that reinforces LLM-generated contextual embeddings with a sequential RNN architecture. The embeddings reprocessing step aims to reinforce sequential semantics and strengthen order-aware dependencies inherent in source code.
We evaluate the proposed hybrid models on both benchmark and real-world coding datasets. The experimental results show that the RoBERTa-BiGRU and CodeBERT-GRU models achieved accuracies of 66.40\% and 66.03\%, respectively, on the defect detection benchmark dataset, representing improvements of approximately 5.35\% ($\uparrow$) and 3.95\% ($\uparrow$) over the standalone RoBERTa and CodeBERT models. Furthermore, the CodeT5-GRU and CodeT5$^+$-BiGRU models achieved accuracies of 67.90\% and 67.79\%, respectively, surpassing their base models and outperforming RoBERTa-BiGRU and CodeBERT-GRU by a notable margin. In addition, CodeT5-GRU model attains weighted ($\psi$) and macro ($\mu$) F1-scores of 67.18\% and 67.00\%, respectively, on the same dataset. Extensive experiments across three real-world datasets (collected from AOJ\footnote{\href{https://onlinejudge.u-aizu.ac.jp/home}{Aizu Online Judge}}) further demonstrate consistent and statistically significant improvements over standalone LLMs. Overall, our findings indicate that reprocessing contextual embeddings with RNN architectures enhances code understanding performance in LLM-based models.
\end{abstract}


\section{Introduction}
\IEEEPARstart{T}he increasing complexity of real-world problems in information technology requires a deep understanding of programming code. Programmers often spend a considerable amount of time searching, collecting, and organizing the necessary information to comprehend and work effectively with code \cite{green1980problems, ko2007information, visser1987strategies, ko2006exploratory, latoza2006maintaining, maalej2014comprehension, meyer2017work, piorkowski2016foraging}. The process is challenging, as it requires the assimilation of various types of knowledge, including code semantics, algorithmic logic, API intricacies, and domain-specific concepts. This information is typically scattered across multiple sources, adding to the difficulty of code comprehension, particularly for novice programmers. An essential aspect of software development is code quality, which encompasses attributes such as structure, readability, and the quality of individual statements. Numerous studies have analyzed static code quality, especially in student-generated code \cite{breuker2011measuring, izu2023exploring, ostlund2023s}, highlighting common programming errors \cite{brown2014investigating} and exploring the semantic properties of code \cite{de2018understanding}. These insights into code quality are critical for improving how developers write and understand code.

In recent years, Large Language Models (LLMs) have become integral to software development, particularly in the area of automated code generation. LLMs such as GPT, Codestral, Copilot, CodeLlama, DeepSeek Coder, CodeT5, and CodeT5$^+$ leverage large-scale corpora of source code and technical documentation to generate code that aligns with user-provided natural language inputs. As a result, LLM-based tools have gained substantial popularity \cite{github_copilot_2024, open3d_2024, tabnine_2024}, as their ability to generate syntactically and semantically correct code against high-level descriptions has proven invaluable to developers. However, as LLMs become more prevalent, researchers have begun to scrutinize the quality of the code produced by these models. Various benchmarks, such as HumanEval \cite{chen2021evaluating}, MBXP \cite{austin2021program}, CoderEval \cite{yu2024codereval}, and ClassEval \cite{du2024evaluating}, have been developed to assess performance on a range of tasks, from basic function generation to complex, class-level code. Moreover, studies have evaluated key attributes of LLM-generated code, including complexity \cite{liu2024no}, security \cite{pearce2022asleep, perry2023users, siddiq2022empirical}, and maintainability \cite{yeticstiren2023evaluating}. Despite these contributions, many of the existing evaluations rely on controlled datasets that may not fully capture the challenges developers face when using LLMs in real-world software development environments. 

While these tools offer the advantage of automatically generating code for developers with limited experience, they also present challenges. Developers may receive code that is difficult to understand or maintain \cite{imai2022github, ziegler2022productivity}, leading to issues in debugging, maintenance, and extensibility \cite{liang2024large, vaithilingam2022expectation}. However, LLMs hold promise in addressing these challenges by providing on-demand, context-specific information that aids in code comprehension.

In this paper, we propose a hybrid LLM–RNN architecture specifically designed for improved code comprehension. The contextual embeddings produced by the LLM are further processed by an RNN to reduce positional bias and strengthen long-range, order-sensitive representations. The LLM serves as the primary encoder, tokenizing and encoding input code sequences into rich contextual embeddings. These embeddings are passed through a dropout layer to mitigate overfitting before being fed into the RNN. The RNN then refines the representations by capturing additional sequential dependencies in the code, improving the model’s ability to understand structural and logical relationships. Finally, a fully connected (FC) layer maps the RNN outputs to the target class labels, and a Softmax function is applied to produce a probability distribution over classes. The main technical contributions of this work are summarized as follows:

   \begin{enumerate}

\item[(i)] We propose a systematic reprocessing strategy that integrates a sequential RNN architecture with domain-specific LLMs to reinforce their contextual embeddings for improved code comprehension tasks. This embedding reprocessing step enables the model to better capture the structural and sequential dependencies inherent in source code. Using both benchmark and real-world datasets, we evaluate the performance of hybrid models such as RoBERTa-, CodeBERT-, CodeT5-, and CodeT5$^+$-RNN against their respective base models. Experimental results consistently demonstrate that the hybrid models outperform their standalone counterparts.

\item[(ii)] Experimental results reveal that CodeT5-RNN achieved significant improvements in both defect detection and real-world datasets. CodeT5-GRU achieved the highest result for code defect detection tasks, with an accuracy of 67.90\%, which represents an improvement of approximately 3.04\% ($\uparrow$) compared to the stand-alone CodeT5-base model. Furthermore, the findings demonstrate that CodeT5-GRU outperformed other models (e.g., RoBERTa-RNN, CodeBERT-RNN, and CodeT5$^+$-RNN ) tested on the same code defect detection dataset. 


\end{enumerate}




The remainder of this paper is organized as follows: Section \ref{relworks} reviews related works and Section \ref{theback} describes the theoretical background. 
Section \ref{prob_state_motivation} presents the problem statements and motivation. Section \ref{proposed_approach} explains the integration process of LLMs and RNNs. Section \ref{experimental_details} details the datasets, evaluation metrics, implementation specifics, baselines, and results. Section \ref{final_discussion} discusses the findings and addresses the research questions. Finally, Section \ref{conclusion_study} concludes the study.

\section{Related Work} \label{relworks}

Code understanding is the task of automatically interpreting and analyzing code snippets to extract meaningful insights about their functionality, structure, and behavior \cite{sharma2021survey,marino2020critical,gao2023code }. This process helps developers gain a deeper comprehension of what the code does, enabling task like defect detection, optimization, and documentation generation. Recent advancements in code intelligence (CI) have led to innovative approaches aimed at improving code understanding and generation. For instance, Rabin et al. \cite{rabin2021understanding} introduce a model-agnostic method that simplifies input programs to identify key syntactic features essential for CI systems, enhancing interpretability and reliability. In the realm of smart contract development, researchers proposed an LSTM-RNN-based method to generate secure contract templates, demonstrating that utilizing contracts with fewer vulnerabilities improves coding security through lexical feature extraction and effective vectorization \cite{hao2023novel}. Furthermore, efforts in multilingual code classification have employed bidirectional long short-term memory (BiLSTM) models, significantly enhancing code classification accuracy by addressing the complexities inherent in diverse programming languages \cite{rahman2023multilingual}.

In the context of program code understanding, PLBART is a sequence-to-sequence model pre-trained on extensive Java and Python code, capable of various tasks such as code summarization and generation. While it excels in discriminative tasks, its limitations, including training on only two languages, highlight the need for expanded language coverage \cite{ahmad2021unified}. Additionally, CodeXGLUE provides a benchmark dataset encompassing 14 datasets and 10 tasks, facilitating structured research comparisons while pointing out the necessity for larger and more specialized datasets to cover niche areas in programming tasks \cite{lu2021codexglue}. CodeBERT, a bimodal pre-trained model for programming and natural languages, uses a Transformer-based architecture trained on both NL-PL pairs and unimodal data. It achieves state-of-the-art performance on tasks like code search and documentation generation and demonstrates superior knowledge in NL-PL probing in zero-shot settings. Despite its success, limitations like task-specific fine-tuning and data quality highlight areas for improvement, such as expanding training data diversity and reducing fine-tuning dependency through enhanced self-supervised techniques \cite{feng2020codebert}. Furthermore, CodeT5 enhances code understanding through a unified encoder-decoder Transformer model, yet faces challenges related to complex code structures and potential biases in training data \cite{wang2021codet5}. Other studies have explored pedagogical approaches, revealing challenges in programming education that emphasize the need for integrated assessment methods to gauge comprehension effectively \cite{kirk2020high, salac2020if}. The Semantic Code Graph model offers a novel representation of code dependencies to improve software comprehension \cite{borowski2024semantic}, while research evaluating the quality of AI-generated code highlights the ongoing challenges of maintainability and effectiveness in code generation systems \cite{liu2024refining, wong2023natural}.

The integration of LLMs and RNNs has shown significant success across various NLP tasks by leveraging the complementary strengths of both architectures \cite{rahman2024roberta}. However, systematic integration of RNNs with LLMs for code comprehension, extensive hyperparameter tuning tailored to such tasks, and a thorough evaluation against standalone models remain largely unexplored. To address these challenges, this article investigates the following key research questions:

\begin{itemize} \item \textbf{RQ1:}  Does reinforcing contextual embeddings with RNN improve model performance?

\item \textbf{RQ2:} How do hybrid LLM-RNN models perform on benchmark and real-world datasets in terms of code comprehension and debugging efficiency?

\item \textbf{RQ3:} What are the advantages of hybrid models over standalone LLMs in handling long-term dependencies and contextual relationships within complex code structures?

\end{itemize}

\section{Theoretical Background} \label{theback}

In this section, we present the theoretical background of language models and analyze their development dynamics over the years. We describe the mathematical representations of statistical language models, neural language models, and LLMs.

\subsection{Statistical Language Model}

The Statistical Language Model (SLM) \cite{jelinek1998statistical, rosenfeld2000two,chu2024history}, introduced in the 1990s, serves as a mathematical approach to capture the contextual properties of natural language from a probabilistic perspective. The core principle of statistical language modeling lies in calculating the probability of a sentence occurring within a given text. Let $S$ be the sentence \texttt{"I love to play football"}. Let $P(s_i)$ represent the probability of the $i$-th word in the sentence $S$, where $s_1$ corresponds to \texttt{"I"}, $s_2$ to \texttt{"love"}, $s_3$ to \texttt{"to"}, $s_4$ to \texttt{"play"}, and $s_5$ to \texttt{"football"}. The likelihood of the entire sentence $P(S)$ can be calculated using the chain rule of probability: 

\begin{equation}
 \begin{split}
&P(S) = P(s_1, s_2, s_3, s_4, s_5) \\
&P(S)= P(\texttt{I}, \texttt{love}, \texttt{to}, \texttt{play}, \texttt{football}) \\
 &P(\texttt{I}, \texttt{love}, \texttt{to}, \texttt{play}, \texttt{football}) = P(\texttt{I}) ~\cdot~ P( \texttt{love}~|~\texttt{I})~ \cdot ~ \\
 &P(\texttt{to}~|~\texttt{I}, \texttt{love}) ~\cdot ~ P(\texttt{play}~|~\texttt{I}, \texttt{love}, \texttt{to}) ~\cdot \\
 &P(\texttt{football}~|~\texttt{I}, \texttt{love}, \texttt{to}, \texttt{play})
 \end{split}
\end{equation}
For a sentence with $n$ words, the above equation can be generalized as follows:
\begin{equation} \label{equation1}
\begin{split}
&P(s_1^n)=P(s_1) \cdot P(s_2| s_1) \cdot P(s_3 | s_1^2)  \cdots P(s_n | s_1^{n-1}) \\
&=\prod_{i=1}^n P(s_i|s_1^{i-1})
\end{split}
\end{equation}
The Markov Assumption is used to describe whether the probability of a word $P(s_i)$ depends on the preceding words $P(s_1^{i-1})$, as follows:

\begin{equation}\label{equation2}
    P(s_i|s_1^{i-1}) \approx P(s_i|s_{i-1})
\end{equation}
Equation (\ref{equation3}) describes an $n$-gram, where $N$ represents the size of the $n$-gram, such as $N=2$ for a bi-gram, $N=3$ for a tri-gram, and so on

\begin{equation}\label{equation3}
    \prod_{i=1}^n P(s_i|s_1^{i-1}) \approx \prod_{i=1}^n P(s_i|s_{i-N+1}^{i-1})
\end{equation}


Maximum likelihood estimation is applied to calculate the conditional probability of each word in a sentence, allowing us to estimate these probabilities by substituting them with observed frequencies using Eq. (\ref{equation4}), where $\mathcal{C}(.)$ denotes the count of occurrences.

\begin{equation}\label{equation4}
    P(s_i|s_{i-N+1}^{i-1}) = \frac{\mathcal{C}(s_{i-N+1}^{i-1} s_i)}{\mathcal{C}(s_{i-N+1}^{i-1})}
\end{equation}

Cross-entropy ($\mathcal{E}$), as described in Eq. (\ref{equation5}), is used to evaluate the performance of an SLM, with a lower value of $\mathcal{E}$ indicating a better-performing model \cite{allamanis2013mining}.

\begin{equation}\label{equation5}
    \mathcal{E} \approx - \frac{1}{n} \sum_{i=1}^n log_2 P(s_i|s_{i-N+1}^{i-1})
\end{equation}

SLMs struggle to handle longer sequences. For instance, with 2,000 different words, there are $2000^n$ possible sequences of length $n$, leading to significant storage limitations as $n$ increases \cite{chu2024history}.

\subsection{Neural Language Model}
Neural Language Models (NLMs) \cite{chung2024scaling, chu2024history,kombrink2011recurrent} use neural networks to predict the probabilities of words in sequences. They effectively handle longer sequences and overcome the limitations of SLMs. RNNs \cite{rahman2023multilingual,rahman2020source,rahman2021bidirectional}, a type of NLM, can process dependent sequential data through their recurrent structure. The following equations describe the RNN model.

\begin{equation}    \emph{\textbf{h}}_t=a(\emph{\textbf{P}}_{(h)}\emph{\textbf{x}}_t + \emph{\textbf{Q}}_{(h)}\emph{\textbf{h}}_{t-1})
\end{equation}
\begin{equation}
    \acute{\emph{\textbf{y}}}_t=o(\emph{\textbf{R}}_{(y)}\emph{\textbf{h}}_t)
\end{equation}

Within the above structure, $\emph{\textbf{x}}_t \in \mathbb{R}^m $ denotes input/word vector, $\emph{\textbf{h}}_t \in \mathbb{R}^n $ denotes hidden layer matrix, and $\acute{\emph{\textbf{y}}}_t \in \mathbb{R}^n$ denotes output vector. The function $a(.)$ represents a non-linearity, and the output of the hidden layer $\emph{\textbf{h}}_t$ is obtained by applying non-linear activation functions, such as $tanh$ or $sigmoid$. Next, the third layer, or output layer, tries to predict the next word by using the Softmax, $o(.)$, activation function, based on the output from the hidden layer.
The matrices $\emph{\textbf{P}}$, $\emph{\textbf{Q}}$, and $\emph{\textbf{R}}$ can be represented  as $\emph{\textbf{P}} \in \mathbb{R}^{m \times n}$, $\emph{\textbf{Q}} \in \mathbb{R}^{n \times n}$, and $\emph{\textbf{R}} \in \mathbb{R}^{n \times n}$, respectively. LSTM, a variant of RNNs, is designed to address the gradient vanishing and exploding issues commonly associated with RNNs. Its gated architecture enables it to handle long-range dependencies in data sequences effectively \cite{rahman2023multilingual}. The LSTM structure can be described as follows.

\begin{equation} \label{equation_lstm}
\begin{split}
&\acute{\emph{\textbf{c}}}_t=a(\emph{\textbf{P}}_{(c)}\emph{\textbf{x}}_t + \emph{\textbf{Q}}_{(c)}\emph{\textbf{h}}_{t-1} + \emph{\textbf{b}}_{(c)}) \\
&\emph{\textbf{f}}_t=\sigma(\emph{\textbf{P}}_{(f)}\emph{\textbf{x}}_t + \emph{\textbf{Q}}_{(f)}\emph{\textbf{h}}_{t-1} + \emph{\textbf{b}}_{(f)})\\
&\emph{\textbf{i}}_t=\sigma(\emph{\textbf{P}}_{(i)}\emph{\textbf{x}}_t + \emph{\textbf{Q}}_{(i)}\emph{\textbf{h}}_{t-1} + \emph{\textbf{b}}_{(i)})\\
&\emph{\textbf{c}}_t=\emph{\textbf{i}}_t 	\odot \acute{\emph{\textbf{c}}}_t + \emph{\textbf{f}}_t \odot \emph{\textbf{c}}_{t-1}\\
&\emph{\textbf{o}}_t=\sigma(\emph{\textbf{P}}_{(o)}\emph{\textbf{x}}_t + \emph{\textbf{Q}}_{(o)}\emph{\textbf{h}}_{t-1} + \emph{\textbf{b}}_{(o)})\\
&\emph{\textbf{h}}_t=\emph{\textbf{o}}_t 	\odot a(\emph{\textbf{c}}_t)
\end{split}
\end{equation}

Within the model structure, $\emph{\textbf{c}}_t \in \mathbb{R}^m $ denotes the state vector. The terms $\emph{\textbf{i}}_t$, $\emph{\textbf{o}}_t$, $\emph{\textbf{c}}_t$, and $\emph{\textbf{f}}_t$  represent the input, output, cell state, and forget gates, respectively. The candidate state, $\acute{\emph{\textbf{c}}}_t \in \mathbb{R}^m$, is generated by a nonlinear function, where $a(.)$ denotes the $tanh(.)$ function, and $\sigma(.)$ represents the sigmoid function. The matrices are defined as follows.  $\emph{\textbf{P}}_{(c)}, \emph{\textbf{P}}_{(f)}, \emph{\textbf{P}}_{(i)}, \emph{\textbf{P}}_{(o)} \in \mathbb{R}^{m \times n}$, $\emph{\textbf{Q}}_{(c)}, \emph{\textbf{Q}}_{(f)}, \emph{\textbf{Q}}_{(i)}, \emph{\textbf{Q}}_{(o)} \in \mathbb{R}^{n \times n}$, and $\emph{\textbf{b}}_{(c)}, \emph{\textbf{b}}_{(f)}, \emph{\textbf{b}}_{(i)}, \emph{\textbf{b}}_{(o)} \in \mathbb{R}^n$. 

The bidirectional LSTM (BiLSTM) \cite{bilstm650093} enhances the performance of the standard LSTM by introducing a backward hidden layer. While the forward hidden layer  ($\overrightarrow{h}$) processes data starting from the first token, the backward hidden layer ($\overleftarrow{h}$) begins processing from the last token. The BiLSTM structure is described as follows.

 \begin{equation} \label{equation_bilstm}
     \begin{split}     &\overrightarrow{\emph{\textbf{h}}}_t=a(\emph{\textbf{P}}_{(fw)}\emph{\textbf{x}}_t + \emph{\textbf{Q}}_{(fw)}\overrightarrow{\emph{\textbf{h}}}_{t-1} + \emph{\textbf{b}}_{(fw)})\\     &\overleftarrow{\emph{\textbf{h}}}_t=a(\emph{\textbf{P}}_{(bw)}\emph{\textbf{x}}_t + \emph{\textbf{Q}}_{(bw)}\overleftarrow{\emph{\textbf{h}}}_{t+1} + \emph{\textbf{b}}_{(bw)})
     \end{split}
 \end{equation} 

  Within the model structure, $\emph{\textbf{P}}_{(fw)},~ \emph{\textbf{P}}_{(bw)} \in \mathbb{R}^{m \times n}$, ~ $\emph{\textbf{Q}}_{(fw)},~ \emph{\textbf{Q}}_{(bw)} \in \mathbb{R}^{n \times n}$, and $\emph{\textbf{b}}_{(fw)},~ \emph{\textbf{b}}_{(bw)} \in \mathbb{R}^{n}$ represent the weight metrics and bias vectors. The forward and backward hidden states ($\overrightarrow{h}$ and $\overleftarrow{h}$) are concatenated and fed into the output layer. 

 \begin{equation} \label{equation_concatenate}
   \emph{\textbf{h}}_t= \overrightarrow{\emph{\textbf{h}}}_t \oplus \overleftarrow{\emph{\textbf{h}}}_t
 \end{equation}

 where  $\emph{\textbf{h}}_{t} \in \mathbb{R}^{m \times 2n}$ represents the weight matrix of the hidden state. The final output, computed at the output layer ($q$: number of outputs), is described as follows. 
 
 \begin{equation}
     \emph{\textbf{o}}_t=o(\emph{\textbf{h}}_t \emph{\textbf{R}}_{(o)} + \emph{\textbf{b}}_{(o)})
 \end{equation} 

where $\emph{\textbf{R}}_{(o)} \in \mathbb{R}^{2n \times q}$ and $\emph{\textbf{b}}_{(o)} \in \mathbb{R}^{q}$ represent weight metrics and bios vector, respectively. The function $o(.)$ denotes the $softmax()$ function, which generates the probability distribution at the output layer.

\subsection{Large Language Model}

The effectiveness of LLMs relies heavily on their extensive model parameters, large and diverse datasets, and substantial computational resources during training \cite{zhao2023survey, jiang2024survey}. Continuous advancements in LLM architectures have improved performance across a wide array of downstream tasks \cite{zhao2023survey}. Given the similar Transformer architecture in LLMs, code-specific LLMs are trained on large-scale unlabeled source code corpora, while general-purpose LLMs, such as ChatGPT \cite{rahman2023chatgpt }, are pre-trained on large-scale datasets of both code and text. Based on their architecture, LLMs are classified into three main categories: encoder-only models, decoder-only models, and encoder-decoder models. 
The Transformer components, including Multi-Head Self-Attention (MHSA), Masked MHSA, Position-wise Feed-Forward Neural Network (PFNN), Residual Connections with Normalization, and Positional Encoding, are briefly described below. The MHSA mechanism is incorporated into each Transformer layer to capture the latent semantic meaning within a sequence across $f$ distinct representation spaces. The MHSA can be represented as follows. 

\begin{equation}
    \emph{\textbf{f}}^{(l)}=\text{MHSA}(Q, K, V)=\text{concat}\{\text{Head}_i\}_{i=1}^f \emph{\textbf{W}}^O
\end{equation}

\begin{equation}
    \text{Head}_i=\text{Attention}\Bigl(\underbrace{\emph{\textbf{F}}^{(l-1)}\emph{\textbf{W}}^Q_i}_Q, \underbrace{\emph{\textbf{F}}^{(l-1)}\emph{\textbf{W}}^K_i}_K, \underbrace{\emph{\textbf{F}}^{(l-1)}\emph{\textbf{W}}^V_i}_V\Bigr)
\end{equation}

\begin{equation}
    \text{Attention}(\emph{\textbf{Q}}, \emph{\textbf{K}}, \emph{\textbf{V}})=\text{softmax}\left(\frac{\emph{\textbf{Q}}\emph{\textbf{K}}^T}{\sqrt{d_{model}/f}}\right)\emph{\textbf{V}}
\end{equation}

Where $\emph{\textbf{F}}^{(l-1)} \in \mathbb{R}^{n \times d_{model}}$ denotes the input to the $l^{th}$ Transformer layer, and  $\emph{\textbf{f}}^{(l)}$ represents the output of MHSA sub-layer. The parameters $f$ and $d_{model}$ denote the number of attention heads and the model dimension, respectively. The projections $\emph{\textbf{W}}^Q_i$, $\emph{\textbf{W}}^K_i$, $\emph{\textbf{W}}^V_i$, $\emph{\textbf{W}}^O_i \in \mathbb{R}^{d_{model} \times d_{model}/f}$ denote the transformation parameters for each attention head $Head_i$.The scaling factor $\sqrt{d_{model}/f}$ is used to scale the dot product of \emph{\textbf{Q}} and \emph{\textbf{K}}, preventing excessively large inner product values. In addition to MHSA, there are two additional attention mechanisms: Masked MHSA and Cross-Layer MHSA. Masked MHSA can be described as follows:
\begin{equation}
\begin{split}    
    &\text{Attention}(\emph{\textbf{Q}}, \emph{\textbf{K}}, \emph{\textbf{V}})=\text{softmax}\left(\frac{\emph{\textbf{Q}}\emph{\textbf{K}}^T}{\sqrt{d_{model}/f}} + \emph{\textbf{M}}_{mask} \right)\emph{\textbf{V}} \\
    &\emph{\textbf{M}}_{mask}=\Bigl(m_{ij}\Bigr)_{n \times n}=\Bigl(\mathbb{I}(i\geq j)\Bigr)_{n \times n}=\Biggl\{\begin{array}{cc}
        0 &  \text{for}~ i \geq j \\
        -\infty &  \text{otherwise}
    \end{array}
\end{split}    
\end{equation}

Masked MHSA is often referred to as causal or autoregressive attention \cite{lin2022survey}. In Cross-Layer MHSA, the \emph{\textbf{K}} and \emph{\textbf{V}} are projected from the encoder outputs, while the \emph{\textbf{Q}} are derived from the previous layer. The PFNN is applied within each Transformer layer to independently refine sequence embeddings at each position $i$, enabling the encoding of more complex feature representations. It consists of two linear transformations with a ReLU activation function in between.

\begin{equation}
    \begin{split}
        &\text{PFNN}(f^{(l)})=\left( \text{concat} \Bigl\{\text{FNN}(f^{(l)}_i)^T\Bigr\}^n_{i=1} \right)^T \\
        &\text{FNN}(f^{(l)}_i)=\text{ReLU}\left(f^{(l)}_i\emph{\textbf{W}}^{(1)} + \emph{\textbf{b}}^{(1)}\right)\emph{\textbf{W}}^{(2)}+\emph{\textbf{b}}^{(2)}
    \end{split}
\end{equation}

Where $\emph{\textbf{W}}^{(1)}, (\emph{\textbf{W}}^{(2)})^T \in \mathbb{R}^{d_{model} \times 4d_{model}}$ represent the projection matrices, while $\emph{\textbf{b}}^{(1)}, \emph{\textbf{b}}^{(2)}$ denote the corresponding bias vectors, all of which are learned during training. The Transformer model includes a residual connection to mitigate gradient vanishing and exploding issues, followed by layer normalization \cite{jiang2024survey}. Another key component of the Transformer is Positional Encoding, which enables the model to capture the positional information of each token. The positional embedding for a token is defined as follows.

\begin{equation}
    \begin{split}
        &\emph{\textbf{P}}_{pos,2i}=sin\left(\frac{pos}{10000^{2i/d_{model}}} \right)\\
        &\emph{\textbf{P}}_{pos,2i+1}=cos\left(\frac{pos}{10000^{2i/d_{model}}} \right)
    \end{split}
\end{equation}

Where $d_{model}$ denotes the model's dimensionality, while  $2i$ and $2i+1$ correspond to the specific dimensions of the positional embedding.

 \begin{figure*} [h]
	\centering
		\includegraphics[width=1\linewidth]{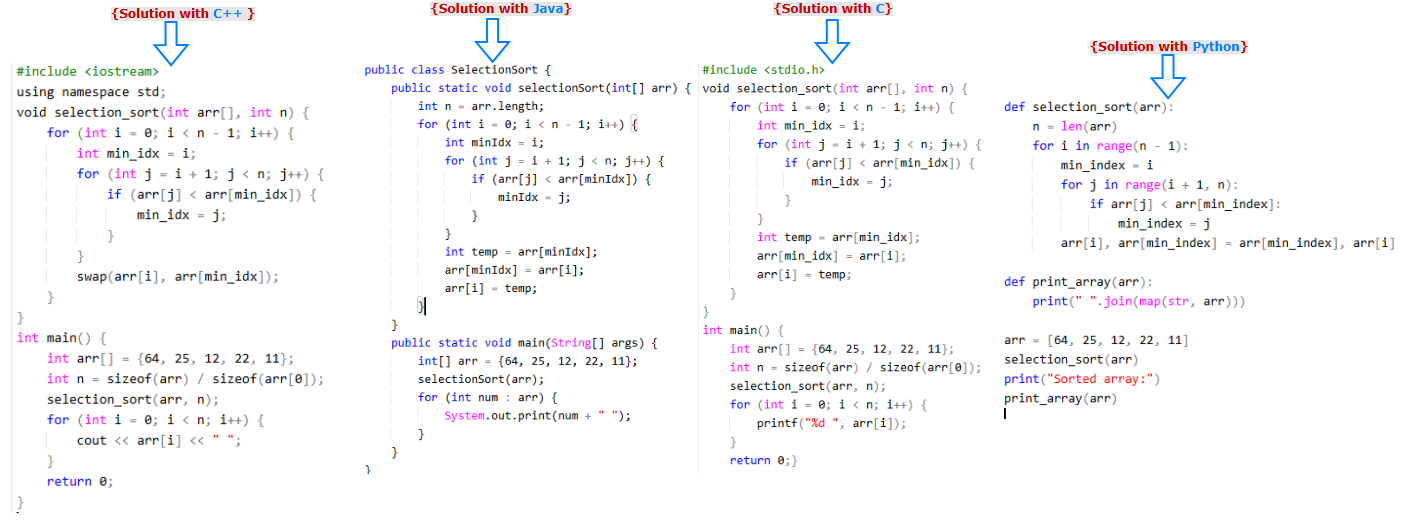}
		\caption {Solution code for the selection sort algorithm implemented in various programming languages (e.g., C++, Java, C, and Python).}
		\label{basic_bilstm_motivation}
\end{figure*}

\section{Problem Statement and Motivation} \label{prob_state_motivation}

Understanding LLM-generated code has become increasingly challenging for programmers, as code generation often varies significantly across models. The challenge is further amplified when the same problem is solved in multiple programming languages. In addition, unlike natural language, programming languages exhibit substantial variation in syntax, structural conventions, and implementation styles across languages, developers, and generative LLMs, resulting in increased complexity and heterogeneity in generated and accumulated solutions.
Figure \ref{basic_bilstm_motivation} illustrates examples of implementations of a selection sort problem, highlighting the diversity in syntax and structure based on specific problem descriptions and constraints\footnote{These example codes are collected from the AOJ platform for problem description  \url{https://onlinejudge.u-aizu.ac.jp/courses/lesson/1/ALDS1/2/ALDS1\_2\_B}.}.These variations demonstrate the inherent challenges in modeling structural and sequential dependencies for robust code understanding.

Recently, domain-specific LLMs, such as CodeBERT, CodeT5, CodeLlama, and GraphCodeBERT, have made notable progress in code understanding \cite{jiang2024survey}. However, despite these advancements, LLMs still achieve suboptimal performance on tasks like defect detection, clone detection, and classification across various benchmark and real-world datasets \cite{zhou2019devign}. These limitations suggest that contextual embeddings produced by LLMs may not fully capture fine-grained structural and long-range sequential dependencies inherent in source code.
Therefore, enhancing the code comprehension capability of LLMs remains a critical research problem. Motivated by this challenge, \textit{we investigate whether reinforcing the contextual embeddings of LLMs with RNN architectures, along with careful hyperparameter optimization, can improve the modeling of structural and order-sensitive dependencies in source code.} 

\section{Methodologies for Reinforcing Contextual Embeddings}\label{proposed_approach}

\begin{figure} [h]
	\centering
		\includegraphics[width=\linewidth]{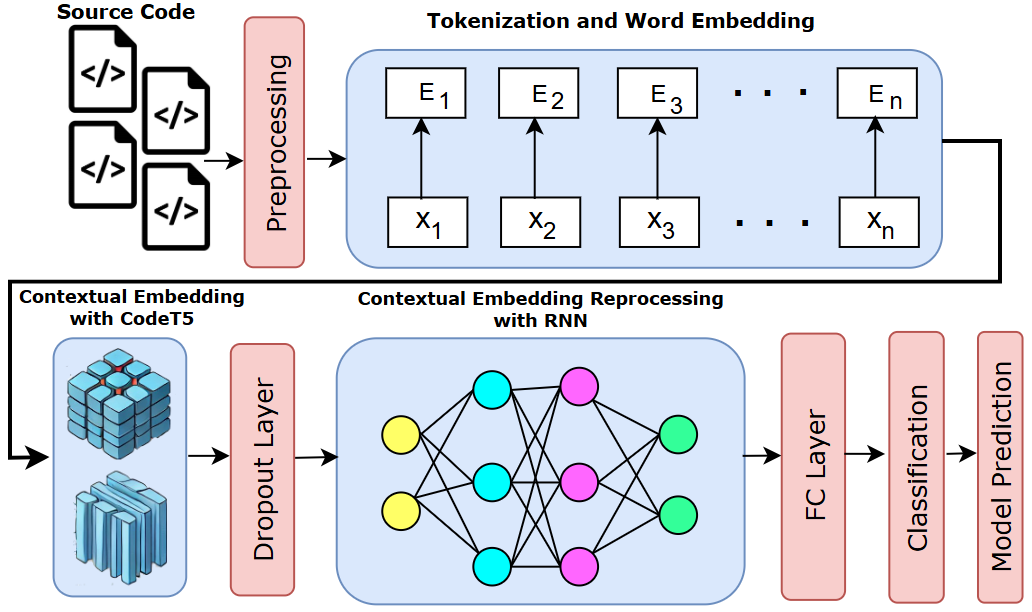}
		\caption {Architecture of the hybrid CodeT5 and RNN model}
		\label{model_structure1}
\end{figure}

This section describes the contextual embedding reinforcing process. Figure \ref{model_structure1} illustrates an overall architecture of the CodeT5-RNN model. Given a sequence of input tokens \( X = \{x_1, x_2, \dots, x_n\} \), the pre-trained CodeT5 model serves as an encoder, transforming the input tokens into embeddings \( E = \{e_1, e_2, \dots, e_n\} \) through $E = \text{CodeT5}(X).$ These embeddings are passed to an RNN layer to capture long-range dependencies, resulting in hidden states \( H = \{h_1, h_2, \dots, h_n\} \) through $H = \text{RNN}(E).$ To enhance generalization and reduce overfitting, a dropout layer is applied to the RNN outputs, producing $H^{\prime} = \text{Dropout}(H),$ where \( H^{\prime} \sim \text{Bernoulli}(p) \) with \( p \) being the dropout probability. The modified outputs \( H^{\prime} \) are then fed into an FC layer, which maps the features into a lower-dimensional space through $Z = W_f \cdot H^{\prime} + b_f,$ where \( W_f \) and \( b_f \) are the weight matrix and bias vector, respectively. The final classification probabilities are obtained using a softmax function, $P(y | X) = \text{Softmax}(Z),$ where $P(y | X) = \frac{\exp(z_i)}{\sum_{j=1}^C \exp(z_j)},$ with \( C \) being the number of classes. The predicted class \( \hat{y} \) is determined as $ \hat{y} = \arg\max_i P(y = i | X).$ This workflow is outlined in Algorithm \ref{algo_code_classification}, with detailed explanations of each component provided in the following sub-sections.

 \begin{algorithm} [h]
    \caption{CodeT5-RNN framework for code understanding.}\label{algo_code_classification}
    \begin{algorithmic}[1]

\STATE  \textbf{Input:} Source codes and code snippets 
\STATE  \textbf{Output:} Predicted class \( \hat{y} \)

\STATE \textbf{Data Prerocessing:} Rigorous data processing is conducted, particularly for real-world datasets. For instance, duplicate solution codes and irrelevant elements are removed. To ensure consistency, cross-validation is performed on the solution codes after the removal of irrelevant elements. Finally, the processed real-world datasets are generated.

\STATE \textbf{Tokenization:} Convert code into tokens (token IDs) using the tokenizer, with attention to padding and truncation. Sequence of input tokens \( X = \{x_1, x_2, \dots, x_n\} \)

 \STATE \textbf{CodeT5 Encoding:} Use the pre-trained CodeT5 model to generate semantic embeddings: \( E = \text{CodeT5}(X), \) where \( E = \{e_1, e_2, \dots, e_n\} \).

\STATE \textbf{Dropout:} Apply dropout to mitigate overfitting and improve model generalization.

\STATE \textbf{Reprocessing of Contextual Embeddings with RNN:} Pass the embeddings \( E \) through an RNN layer to capture long-range dependencies and enhance feature representation: \( H = \text{RNN}(E), \) where \( H = \{h_1, h_2, \dots, h_n\} \).

\STATE \textbf{Dropout:} Apply dropout to the RNN outputs to reduce overfitting: \( H^{\prime} = \text{Dropout}(H), \).

\STATE \textbf{FC Layer:} Transform the dropout-regularized outputs into a lower-dimensional space for further processing.

\STATE \textbf{Classification Layer:} Apply a classification layer to predict the class label for each solution code or snippet.

\STATE \textbf{Prediction:} Determine the predicted class by selecting the one with the highest probability: \( \hat{y} = \arg\max_i P(y = i | X). \)

\STATE \textbf{Return:} Predicted class \( \hat{y} \)
    \end{algorithmic}
\end{algorithm}

\subsection{Contextual Embeddings Generation with CodeT5}
CodeT5 \cite{wang2021codet5}, a pre-trained encoder-decoder Transformer model tailored for code-related tasks. CodeT5 is built on the Transformer architecture, consisting of 12 layers each for both the encoder and decoder, totaling 24 layers in the model architecture (12 encoder layers + 12 decoder layers), with a hidden size of 768 and 12 attention heads. This robust architecture allows the CodeT5 model to understand the latent semantic, dependencies, and structure of code. CodeT5 is pre-trained on a large-scale corpus of 8.35 million code snippets from various programming languages \cite{husain2019codesearchnet, CodeRL2022}, including Python, Java, JavaScript, Ruby, PHP, and Go, sourced from open-source repositories like GitHub. This diverse and extensive training data enables CodeT5 to generalize well across different programming languages and adapt to a variety of code-related tasks. A key innovation in CodeT5 is its identifier-aware pre-training task, where the model is trained to recognize and recover masked identifiers (such as variable and function names). This identifier awareness is crucial, as it enables CodeT5 to capture the semantic richness of code structures, deepening its understanding of interrelationships within code snippets. Furthermore, CodeT5 supports multi-task learning within a unified framework, seamlessly managing both code understanding tasks (e.g., defect detection and code clone detection) and code generation tasks (e.g., PL-NL, NL-PL, and code-to-code translation). 
In particular, the CodeT5 encoder model is used in this study.

CodeT5 uses \textit{Byte Pair Encoding (BPE)} for tokenization, ensuring efficient representation of the input while minimizing the out-of-vocabulary (OOV) issue. Let \( X \) represent the raw input code or text, and the tokenization process can be expressed as \( T = \text{BPE}(X) = \{t_1, t_2, \dots, t_n\} \), where \( T \) is the sequence of tokens. Each token \( t_i \) is mapped to an \textbf{Input ID} \( \text{id}_i \in \mathbb{Z}^+ \), a \textbf{Token type ID} \( \text{tt}_i \in \{0, 1\} \), and an \textbf{Attention mask} \( \text{am}_i \in \{0, 1\} \), enabling focused self-attention. CodeT5 employs a \textit{denoising objective} based on span masking, where the masked sequence \( T' = \{t_1, \dots, [\text{MASK}], \dots, t_n\} \) is reconstructed to minimize the loss \( \mathcal{L}_\text{recon} = - \sum_{i \in \text{mask}} \log P(t_i \mid T') \), with \( P(t_i \mid T') \) representing the probability of reconstructing the masked token. Using a Transformer architecture, token embeddings \( E(T') = \{e_1, e_2, \dots, e_n\} \) are derived, where \( e_i = \text{Embed}(t_i) \). These embeddings are processed by the Transformer encoder to produce contextual representations \( H(T') = \text{TransformerEncoder}(E(T')) = \{h_1, h_2, \dots, h_n\} \), with \( h_i \) being the contextual embedding for \( t_i \). The overall objective combines the span reconstruction loss with auxiliary tasks as \( \mathcal{L} = \mathcal{L}_\text{recon} + \mathcal{L}_\text{aux} \), where \( \mathcal{L}_\text{aux} \) includes tasks such as code completion. The pipeline can be summarized as \( T \xrightarrow{\text{BPE}} T' \xrightarrow{\text{Embedding}} E(T') \xrightarrow{\text{Transformer Encoder}} H(T') \xrightarrow{\text{RNN \& Clssifier}} \hat{y} \), enabling robust contextual learning and efficient handling of code and text sequences.

\subsection{Reinforcing Contextual Embeddings with RNN }





The CodeT5-RNN highlights the effectiveness of RNN models in capturing rich contextual details, establishing them as a popular choice for sequential data analysis tasks due to their enhanced performance and resilience. The output embeddings from the final layer \( L \) of the CodeT5 model are represented as a sequence \( H^{(L)} = \{h_1^{(L)}, h_2^{(L)}, \dots, h_n^{(L)}\} \), where \( h_i^{(L)} \in \mathbb{R}^d \) denotes the \( i \)-th embedding in the sequence, and \( d \) is the dimensionality of the embeddings. To reduce overfitting, a dropout operation is applied to these embeddings, resulting in \( h_i^{\text{drop}} = \text{Dropout}(h_i^{(L)}) \), where \( h_i^{\text{drop}} \in \mathbb{R}^d \). To align the dimensionality of the CodeT5 output embeddings with the input requirements of the RNN, a linear transformation is applied to each embedding: \( z_i = W_{\text{linear}} h_i^{\text{drop}} + b_{\text{linear}} \), where \( z_i \in \mathbb{R}^{d_{\text{RNN}}} \), \( W_{\text{linear}} \in \mathbb{R}^{d_{\text{RNN}} \times d} \) is the weight matrix, and \( b_{\text{linear}} \in \mathbb{R}^{d_{\text{RNN}}} \) is the bias vector. The sequence of transformed embeddings \( \{z_1, z_2, \dots, z_n\} \) is then reprocessed by the RNN, which computes the hidden states sequentially: \( h_i^{\text{RNN}} = \text{RNN}(z_i, h_{i-1}^{\text{RNN}}) \), where \( h_i^{\text{RNN}} \in \mathbb{R}^{d_{\text{RNN}}} \) is the \( i \)-th hidden state, and \( h_{i-1}^{\text{RNN}} \) is the hidden state from the previous time step. The final output sequence of the RNN is given by \( H_{\text{RNN}} = \{ h_1^{\text{RNN}}, h_2^{\text{RNN}}, \dots, h_n^{\text{RNN}} \} \), which combines the contextual information from CodeT5 with the sequential dependencies modeled by the RNN.

\subsection{FC and Classification Layers}

A dropout layer is applied to \( H_{\text{RNN}} \), $H^{\prime} = \text{Dropout}(H_{\text{RNN}}),$ to mitigate overfitting, followed by an FC layer that maps the RNN outputs to class logits:

\begin{equation} 
    Z_i = \text{ReLU}(W_{\text{dense}} H^{\prime} + b_{\text{dense}}) 
\end{equation}

Finally, a softmax function is applied to the dense layer output, producing a probability distribution over classes:

\begin{equation} 
    P(y_i | X) = \text{Softmax}(W_o Z_i + b_o) 
\end{equation}


\section{Experimental Settings}

In this section, we describe the overall experimental setup, including the computational environment, hyperparameter settings, evaluation metrics, datasets, and baseline models leveraged for the experiments.

\subsection{Implementation Details}

The experiments are conducted within the operating system environment of Ubuntu 22.04.4 LTS 64-bit. The hardware specifications are detailed as follows: Processor: AMD Ryzen 9 3950X 16-core processor with 32 threads, RAM: 64GB, Graphics: NVIDIA GeForce RTX 3090/PCIe/SSE2, Graphics Memory: 24GB, and Disk Capacity: 500GB.

\subsection{Hyperparameters}

The performance of LLMs heavily relies on selecting suitable hyperparameters. In this paper, we conduct comprehensive experiments using various hyperparameter sets to investigate hybrid model performance in code understanding tasks. Table \ref{hyperpameters} presents the hyperparameters used for model fine-tuning. Each RNN model is paired with an LLM to construct a hybrid model. Note that the number of RNN hidden units (RHUs), denoted by $h$, is doubled ($2 \times h$) when using BiLSTM and BiGRU, due to their bidirectional data processing capabilities ($\overrightarrow{h}$ and $\overleftarrow{h}$). Categorical cross-entropy is used to calculate the training loss, defined as follows:

\begin{equation} \label{lossfunction}
     \mathcal{L}(g)=-\sum_{j=1}^K u_j ~log(\bar{u_j})
 \end{equation}
where $g$ and $K$ represent the model parameter and the number of classes, respectively, while $u_j$ and $\bar{u_j}$ denote the true and predicted labels, respectively, for the $j^{th}$ sample.

\begin{table}[h]
\caption{Selected hyperparameters for the experiments}
\begin{center}
\begin{tabular}{p{3cm}|p{5cm}}
\hline \hline
\textbf{Parameter-Name} & \textbf{Values} \\
\hline \hline
Pre-trained LLMs & 	{\RaggedRight{RoBERTa (\texttt{roberta-base} \cite{liu2019roberta}), CodeBERT (\texttt{microsoft/ codebert-base} \cite{feng2020codebert}), CodeT5 (\texttt{Salesforce/ codet5-base} \cite{wang2021codet5}), CodeT5$^+$ (\texttt{Salesforce/codet5p-220m \cite{wang2023codet5plus}}) }}  \\ \hline
RNNs & LSTM, BiLSTM, GRU, BiGRU  \\ \hline
Optimizer ($\Delta$) & AdamW, NAdam, RMSprop \\ \hline
Loss function ($\mathcal{L}$)  & \RaggedRight{Categorical Cross Entropy (\texttt{cross\_entropy})} \\ \hline

Epochs ($epoch$)& 5 \\ \hline
Dropout ($d$)& 0.1, 0.2 \\ \hline
Learning rates ($l$) & $1e^4$, $1e^5$, $2e^5$, $1e^6$\\ \hline
RNN Hidden units (RHUs) ($h$) & 128, 256, 512\\ \hline
Datasets & code\_x\_glue\_cc\_defect\_detection, SearchAlg, SearchSortAlg, SearchSortGTAlg \cite{zhou2019devign,rahman2023multilingual} \\ \hline
Train/Val/Test & 80\%/10\%/10\%\\ \hline
\end{tabular}
\label{hyperpameters}
\end{center}
\end{table}

\subsection{Evaluation Metrics}


The performance of the proposed models is evaluated on code comprehension, defect detection, code classification, and algorithm identification tasks using standard metrics, including accuracy (A), weighted ($\psi$) and macro ($\mu$) precision (P), recall (R), and F1-score (F1) \cite{rahman2023multilingual, ref19, ref60}.

The metric for accuracy (A) is defined as follows.

 \begin{equation}\label{class_accuracy}
 \emph{\textbf{A}}=\frac{1}{N}\sum^{|K|}_{l=1} \sum_{x:f(x)=l} H(f(x) = \hat{f}(x))
 \end{equation}
 
where $H$ is the function that returns 1 if the class is true and 0, otherwise. $K$ denotes the total number of classes, and $f(x) \in 
K=\{1, 2, 3, \cdots\}$. In this study, the weighted-precision ($\emph{\textbf{P}}_\psi$), recall ($\emph{\textbf{R}}_\psi$), and F1-score ($\emph{\textbf{F1}}_\psi$) are calculated to ensure an unbiased performance evaluation. These metrics are defined as follows. 
 
   \begin{equation}\label{weight_p}
     \emph{\textbf{P}}_\psi= \frac{1}{|N|} \sum_{j=1}^{|K|} \frac{TP_j}{TP_j + FP_j} \times |n_j| = \frac{\sum_{j=1}^{|K|} \emph{\textbf{P}}_j \times |n_j|}{|N|}
 \end{equation}
 
   \begin{equation}\label{weight_r}
     \emph{\textbf{R}}_\psi= \frac{1}{|N|} \sum_{j=1}^{|K|} \frac{TP_j}{TP_j + FN_j} \times |n_j| = \frac{\sum_{j=1}^{|K|} \emph{\textbf{R}}_j \times |n_j|}{|N|}
 \end{equation}

   \begin{equation}\label{weight_f1}
     \emph{\textbf{F1}}_\psi= \frac{1}{|N|} \sum_{j=1}^{|K|} \emph{\textbf{F1}}_j \times |n_j|
 \end{equation}

Here, $|n_j|$ represents the support of the $j^{th}$ class, and $|N|$ denotes the total support,  or the sum of instances across all classes. When the class distribution is imbalanced, the macro-average ($\mu$) is more appropriate. A higher macro F1-score indicates better model performance across individual classes. The macro-precision ($\emph{\textbf{P}}_\mu$), recall ($\emph{\textbf{R}}_\mu$), and F1-score ($\emph{\textbf{F1}}_\mu$) are calculated as follows.
  \begin{equation}\label{macro_p}
     \emph{\textbf{P}}_\mu= \frac{1}{|N|} \sum_{j=1}^{|K|} \frac{TP_j}{TP_j + FP_j} = \frac{\sum_{j=1}^{|K|} \emph{\textbf{P}}_j}{|N|}
 \end{equation}
 
  \begin{equation}\label{macro_r}
     \emph{\textbf{R}}_\mu= \frac{1}{|N|} \sum_{j=1}^{|K|} \frac{TP_j}{TP_j + FN_j} = \frac{\sum_{j=1}^{|K|} \emph{\textbf{R}}_j}{|N|}
 \end{equation}
 
   \begin{equation} \label{macro_f1}
     \emph{\textbf{F1}}_\mu=\frac{ 2 \times \emph{\textbf{P}}_\mu \times \emph{\textbf{R}}_\mu}{\emph{\textbf{P}}_\mu + \emph{\textbf{R}}_\mu}
 \end{equation}

\subsection{Datasets}

Benchmark and real-world datasets are meticulously prepared for the experiments. The defect detection benchmark dataset, sourced from CodeXGLUE \cite{zhou2019devign}, is primarily used to evaluate the model’s effectiveness in identifying code defects. This dataset includes three attributes: $func$, $target$, and $idx$. The attribute $func$ contains the source code, while $target$ is a binary indicator with values 0 or 1, representing the presence or absence of a vulnerability, respectively. The other three datasets, namely SearchAlg, SearchSortAlg, and SearchSortGTAlg, were collected from AOJ \cite{ref35,10811933}, a well-regarded repository of real-world code. AOJ has been recognized as a reliable source, with organizations like Google DeepMind and IBM utilizing its codebase for their CodeNet and Alphacode projects \cite{ref57,ref58}, respectively. The SearchAlg dataset contains source code for 05 different search algorithms. Similarly, SearchSortAlg comprises source code of 15 sorting and search algorithms, while SearchSortGT includes solution codes of 29 different algorithms, such as those for graph and tree structures, as well as searching and sorting algorithms. The data collection and preprocessing steps for real-world datasets were conducted following the methodology outlined in \cite{rahman2023multilingual}. An overview of these datasets is provided in Table \ref{datasets_details}.

\begin{table}[!h]
\caption{Overview of the datasets and their distributions.} \label{datasets_details}
\centering
\begin{tabular}{c|l||p{1cm}|p{.7cm}|p{2.4cm}}
\hline \hline
\textbf{Sl.} & \textbf{Datasets} & \textbf{No. of Codes}  & \textbf{Classes} & \textbf{No. of Train/Val/Test Codes} \\
\hline \hline

1 & Defect Detection & 25400 & 2 & 20320/2540/2540 \\ \hline

2 & SearchAlg & 25994 & 5 & 20795/2599/2600\\ \hline

3 & SearchSortAlg & 80745 & 15 & 64596/8075/8075\\ \hline

4 & SearchSortGTAlg & 119476  & 29  & 95580/11947/11947\\ 
\hline
\end{tabular}
\end{table}

\begin{table*}[!t]
\caption{The best performance ($\mathbf{F1}_\psi$, $\mathbf{P}_\psi$, and $\mathbf{R}_\psi$) achieved by RoBERTa-RNN models on defect detection dataset} \label{roberta_rnn_model_p_r_f1}
\centering
\begin{tabular}{p{1.2cm}|p{1.4cm}||p{.6cm}|p{.65cm}|p{.65cm}|p{.65cm} ||p{.6cm}|p{.65cm}|p{.65cm}|p{.65cm}||p{.6cm}|p{.65cm}|p{.65cm}|p{.65cm}}
\hline \hline

\multirow{2}{*}{\textbf{Model}} & \multirow{2}{*}{\textbf{\makecell{Learning \\Rate ($l$)}}} & \multicolumn{4}{c||} {\textbf{Adam}} & \multicolumn{4}{c||} {\textbf{NAdam}} & \multicolumn{4}{c}{\textbf{RMSprop}} \\ \cline{3-14}

 &  & RHUs & \centering $\mathbf{F1}_\psi$ & \centering $\mathbf{P}_\psi$ &  \makecell{$\mathbf{R}_\psi$}  & RHUs & \centering $\mathbf{F1}_\psi$ & \centering $\mathbf{P}_\psi$ & \makecell{$\mathbf{R}_\psi$}  & RHUs & \centering $\mathbf{F1}_\psi$ & \centering $\mathbf{P}_\psi$ &  \makecell{$\mathbf{R}_\psi$} \\ \hline \hline

\multirow{4}{*}{\makecell{RoBERTa-\\LSTM}} & \multirow{1}{*}{$1e^{-4}$} & 128 & 0.3794 & 0.2922 & 0.5406 &	128 & 0.3794 & 0.2922 & 0.5406 &	128 & 0.3794 & 0.2922 & 0.5406 \\ \cline{2-14}

& \multirow{1}{*}{$1e^{-5}$} & 128 & 0.6372 & 0.6450 & 0.6445 &	512 & 0.6446 & 0.6470 & 0.6482 &  256 & 0.5549 & 0.6526 & 0.6098\\  \cline{2-14}

& \multirow{1}{*}{$1e^{-6}$} & 128 & 0.5967 & 0.5990 & 0.6018 &	512 & 0.5934 & 0.6188 & 0.6138 & 128 & 0.5817 & 0.6212 & 0.6105\\ \cline{2-14}

& \multirow{1}{*}{$2e^{-5}$} & 128 & 0.3794 & 0.2923 & 0.5406 & 128 & 0.3794 & 0.2923 & 0.5406 & 128 & 0.3794 & 0.2923 & 0.5406 \\ \hline


\multirow{4}{*}{\makecell{RoBERTa-\\BiLSTM}} & \multirow{1}{*}{$1e^{-4}$} & 128  & 0.3794 & 0.2923 & 0.5406 &	128 & 0.3794 & 0.2923 & 0.5406 & 128 & 0.3794 & 0.2923 & 0.5406\\  \cline{2-14} 

& \multirow{1}{*}{$1e^{-5}$} & 512 &	0.6413 & 0.6417 & 0.6431 &	128 & 0.3794 & 0.2923 & 0.5406 & 512 &	0.6168 & 0.6652 & 0.6435 \\ \cline{2-14} 

& \multirow{1}{*}{$1e^{-6}$} & 512 & 0.6013 & 0.6100 & 0.6113 &	128 & 0.5895 & 0.6030 & 0.6036 & 256 & 0.5818 & 0.6203 & 0.6102 \\ \cline{2-14}

& \multirow{1}{*}{$2e^{-5}$} & 128 & 0.3794 & 0.2923 & 0.5406 &	128 & 0.3794 & 0.2923 & 0.5406 &	128 & 0.3794 & 0.2923 & 0.5406 \\  \hline

\multirow{4}{*}{\makecell{RoBERTa-\\GRU}} & \multirow{1}{*}{$1e^{-4}$} & 128  & 0.3794 & 0.2923 & 0.5406 &	128 & 0.3794 & 0.2923 & 0.5406 &	128 & 0.3794 & 0.2923 & 0.5406\\ \cline{2-14}

& \multirow{1}{*}{$1e^{-5}$} & 128  & 0.6416 & 0.6423 & 0.6413 & 512 & 0.6530 & 0.6581 & 0.6581& 128 & 0.3794 & 0.2923 & 0.5406\\ \cline{2-14} 

& \multirow{1}{*}{$1e^{-6}$} & 256 & 0.5993 & 0.5995 & 0.5992 & 128 & 0.5976 & 0.6071 & 0.6083 & 256 & 0.5856 & 0.6114 & 0.6072\\ \cline{2-14}

& \multirow{1}{*}{$2e^{-5}$} & 128  & 	0.3794 & 0.2923 & 0.5406 & 128 & 0.3794 & 0.2923 & 0.5406 & 128 & 0.3794 & 0.2923 & 0.5406\\ \hline

\multirow{4}{*}{\makecell{RoBERTa-\\BiGRU}} & \multirow{1}{*}{$1e^{-4}$} & 128  & 0.3794  & 0.2923 & 0.5406 &	128 & 0.3794 & 0.2923 & 0.5406 & 128 &	0.3794 & 0.2923 & 0.5406\\  \cline{2-14} 

& \multirow{1}{*}{$1e^{-5}$} & 128  & 0.6489 & 0.6498 & 0.6512 & 512 & 0.6476 & 0.6773 & 0.6640  & 512 & 0.5912 & 0.6655 & 0.6307 \\ \cline{2-14} 

& \multirow{1}{*}{$1e^{-6}$} & 256 & 0.6016 & 0.6029 & 0.6054 &	256 & 0.6023 & 0.6028 & 0.6051 & 128 & 0.5927 & 0.6034 & 0.6047 \\  \cline{2-14}

& \multirow{1}{*}{$2e^{-5}$} & 128  & 	0.3794 & 0.2923 & 0.5406  & 128 & 0.3794 & 0.2923 & 0.5406 & 128 & 0.3794 & 0.2923 & 0.5406\\ \hline 

\end{tabular}

\end{table*}

\subsection{Baseline}

We compare the proposed models against several strong baseline LLMs, including PLBART \cite{ahmad2021unified}, RoBERTa \cite{zhou2019devign}, CodeT5, CodeBERT+GFSA \cite{choi2023graph}, CodeT5-Small, C-BERT \cite{buratti2020exploring}, CoTexT \cite{phan2021cotext}, CodeBERT \cite{zhou2019devign}, RoBERTa+GFSA \cite{choi2023graph}, and CodeT5$^+$. These models represent state-of-the-art approaches for code understanding tasks. We evaluate the performance of our proposed model against these baselines on the benchmark dataset. Furthermore, we conduct extensive hyperparameter tuning and systematically investigate multiple RNN variants (e.g., LSTM, GRU, BiLSTM, and BiGRU) when integrating recurrent architectures with LLMs.

\section{Experimental Results} \label{experimental_details}

In this section, we present the experimental evaluation of the proposed hybrid models on both benchmark and real-world datasets. Our study examines several base LLMs, including RoBERTa, CodeBERT, CodeT5, and CodeT5$^+$, each integrated with different RNN architectures.

\begin{figure} []
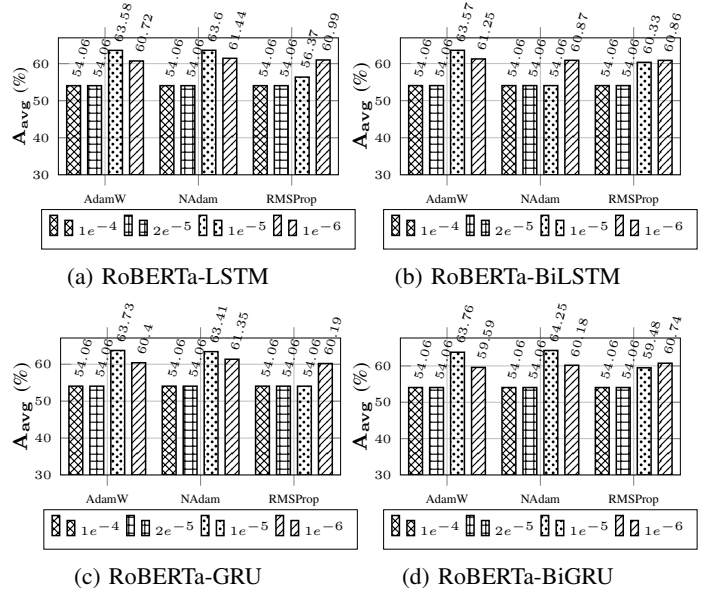

     \centering
     \begin{subfigure}[b]{0.49\linewidth}
         \captionsetup{justification=centering}
         \begin{tikzpicture}[scale=.95]
            \input{roberta_accuracy_LR_0.0001}
        \end{tikzpicture}
        \caption{RoBERTa-LSTM}
         \label{roberta_accuracy_LR_0.0001}
     \end{subfigure}
     \hfill
       \begin{subfigure}[b]{0.49\linewidth}
        \captionsetup{justification=centering}
         \begin{tikzpicture}[scale=.95]
            \input{roberta_accuracy_LR_0.00001}
        \end{tikzpicture}
        \caption{RoBERTa-BiLSTM}
         \label{roberta_accuracy_LR_0.00001}
     \end{subfigure}
     \hfill
     \begin{subfigure}[b]{0.49\linewidth}
        \captionsetup{justification=centering}
         \begin{tikzpicture}[scale=.95]
            \input{roberta_accuracy_LR_0.00002}
        \end{tikzpicture}
       \caption{RoBERTa-GRU}
         \label{roberta_accuracy_LR_0.00002}
     \end{subfigure}
     \hfill     
     \begin{subfigure}[b]{0.49\linewidth}
        \captionsetup{justification=centering}
         \begin{tikzpicture}[scale=.95]
            \input{roberta_accuracy_LR_0.000001}
        \end{tikzpicture}
       \caption{RoBERTa-BiGRU}
         \label{roberta_accuracy_LR_0.000001}
     \end{subfigure}
   \vspace{-5mm}


          \caption{$\mathbf{A_{avg}}$ scores of the RoBERTa-RNN models for code understanding on the defect detection dataset, evaluated with $l=\{1e^{-4}, ~2e^{-5}, ~1e^{-5}, ~1e^{-6}\}$, $\Delta=$\{AdamW, NAdam, RMSprop\}, and $h=\{128, 256, 512\}$.}

        \label{roberta_rnn_accuracy}
\end{figure}

\begin{figure}[]
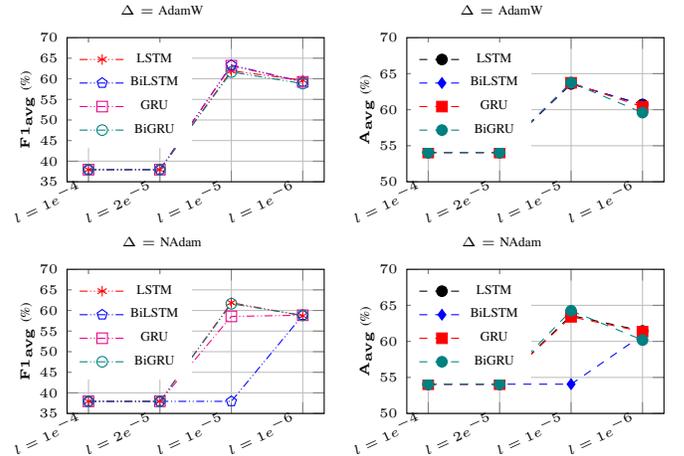

     \centering
     \begin{subfigure}[b]{0.49\linewidth}     
         \captionsetup{justification=centering}
         \begin{tikzpicture}[scale=1]
            \input{roberta_lstm_f1_performance_with_lr}      
            \node[above,font=\tiny] at (current bounding box.north) {$\Delta=$ AdamW};
        \end{tikzpicture}
         \label{roberta_lstm_f1_performance_with_lr}
         \vspace{-5mm}
     \end{subfigure}
     \hfill
       \begin{subfigure}[b]{0.49\linewidth}
        \captionsetup{justification=centering}
         \begin{tikzpicture}[scale=1]
            \input{roberta_lstm_accuracy_performance_with_lr}
        \node[above,font=\tiny] at (current bounding box.north) {$\Delta=$ AdamW};
        \end{tikzpicture}
         \label{roberta_lstm_accuracy_performance_with_lr}
        \vspace{-5mm}
     \end{subfigure}
     \hfill     
        \begin{subfigure}[b]{0.49\linewidth}
        \captionsetup{justification=centering}
         \begin{tikzpicture}[scale=1]
            \input{roberta_lstm_f1_performance_with_lr_nadam}
        \node[above,font=\tiny] at (current bounding box.north) {$\Delta=$ NAdam};
        \end{tikzpicture}
         \label{roberta_lstm_f1_performance_with_lr_nadam}
     \end{subfigure}
     \hfill
            \begin{subfigure}[b]{0.49\linewidth}
        \captionsetup{justification=centering}
         \begin{tikzpicture}[scale=1]
            \input{roberta_lstm_accu_performance_with_lr_nadam}
        \node[above,font=\tiny] at (current bounding box.north) {$\Delta=$ NAdam};
        \end{tikzpicture}
         \label{roberta_lstm_accu_performance_with_lr_nadam}
     \end{subfigure}    
\vspace{-10mm}


          \caption{Comparative analysis of $\mathbf{F1_{avg}}$ and $\mathbf{A_{avg}}$ for RoBERTa-RNN models using $\Delta=$\{AdamW, NAdam\}, $h=\{128, 256, 512\}$, and  $l=\{1e^{-4}, ~2e^{-5}, ~1e^{-5}, ~1e^{-6}\}$.}

        \label{roberta_rnn_f1_a}
\end{figure}

\begin{table*}[!t]
\caption{The best performance ($\mathbf{F1}_\psi$, $\mathbf{P}_\psi$, and $\mathbf{R}_\psi$) achieved by CodeBERT-RNN models on defect detection dataset} \label{codebert_rnn_model_p_r_f1}
\centering
\begin{tabular}{p{1.2cm}|p{1.4cm}||p{.6cm}|p{.65cm}|p{.65cm}|p{.65cm} ||p{.6cm}|p{.65cm}|p{.65cm}|p{.65cm}||p{.6cm}|p{.65cm}|p{.65cm}|p{.65cm}}
\hline \hline

\multirow{2}{*}{\textbf{Model}} & \multirow{2}{*}{\textbf{\makecell{Learning \\Rate ($l$)}}} & \multicolumn{4}{c||} {\textbf{Adam}} & \multicolumn{4}{c||} {\textbf{NAdam}} & \multicolumn{4}{c}{\textbf{RMSprop}} \\ \cline{3-14}

 &  & RHUs & \centering $\mathbf{F1}_\psi$ & \centering $\mathbf{P}_\psi$ &  \makecell{$\mathbf{R}_\psi$}  & RHUs & \centering $\mathbf{F1}_\psi$ & \centering $\mathbf{P}_\psi$ & \makecell{$\mathbf{R}_\psi$}  & RHUs & \centering $\mathbf{F1}_\psi$ & \centering $\mathbf{P}_\psi$ &  \makecell{$\mathbf{R}_\psi$} \\ \hline \hline

\multirow{4}{*}{\makecell{CodeBERT-\\LSTM}} & \multirow{1}{*}{$1e^{-4}$} & 128 & 0.3794 & 0.2923 & 0.5406 &	128 & 0.3794 & 0.2923 & 0.5406 &	128 & 0.3794 & 0.2923 & 0.5406 \\  \cline{2-14} 

& \multirow{1}{*}{$1e^{-5}$} & 256 &	0.6026 & 0.6294 & 0.6226 & 128 & 0.6289 & 0.6499 & 0.6435 &  256 & 0.6285 & 0.6520 & 0.6442 \\ \cline{2-14} 

& \multirow{1}{*}{$1e^{-6}$} & 128 & 0.5876 & 0.6511 & 0.6245 &	128 & 0.5680 & 0.6546 & 0.6164 &  512 & 0.5729 & 0.6543 & 0.6186\\ \cline{2-14}
 
& \multirow{1}{*}{$2e^{-5}$} & 512 & 0.6167 & 0.6200 & 0.6219 & 256 & 0.6513 & 0.6544 & 0.6552 & 128 & 00.5967 & 0.6074 & 0.6083 \\ \hline

\multirow{4}{*}{\makecell{CodeBERT-\\BiLSTM}} & \multirow{1}{*}{$1e^{-4}$} & 128  &	0.3794 & 0.2923 & 0.5406 & 128 &	0.3794 & 0.2923 & 0.5406 & 128 &	0.3794 & 0.2923 & 0.5406\\ \cline{2-14} 

& \multirow{1}{*}{$1e^{-5}$} & 128 &0.6503 & 0.6560 & 0.6559 &	512 & 0.6406 & 0.6641 & 0.6552  & 512 & 0.6316 & 0.6554 & 0.6471 \\  \cline{2-14} 

& \multirow{1}{*}{$1e^{-6}$} & 512 &	0.5811 & 0.6537 & 0.6223 & 512 &	0.5761 & 0.6619 & 0.6223 & 512 &	0.5658 & 0.6597 & 0.6168 \\ \cline{2-14}
  
& \multirow{1}{*}{$2e^{-5}$} & 128 & 0.6420 & 0.6417 & 0.6424 &	 256 & 0.6571 & 0.6571 & 0.6570 & 512 & 0.5861 & 0.6880 & 0.6340  \\  \hline

\multirow{4}{*}{\makecell{CodeBERT-\\GRU}} & \multirow{1}{*}{$1e^{-4}$} & 128  &	0.3794 & 0.2923 & 0.5406 & 128 &	0.3794 & 0.2923 & 0.5406 & 128 &	0.3794 & 0.2923 & 0.5406\\ \cline{2-14}

& \multirow{1}{*}{$1e^{-5}$} & 512  & 0.6235 & 0.6299 & 0.6307 & 256 &  0.6207 & 0.6221 & 0.6241 & 256 & 0.6360 & 0.6607 & 0.6515 \\ \cline{2-14} 

& \multirow{1}{*}{$1e^{-6}$} & 128  & 0.5782 & 0.6475 & 0.6190 & 512 & 0.5728 & 0.6427 & 0.6149 & 128 & 0.5827 & 0.6514 & 0.6223 \\ \cline{2-14}

& \multirow{1}{*}{$2e^{-5}$} & 512  & 	0.6532 & 0.6620 & 0.6603 & 128 & 0.6529 & 0.6630 & 0.6607 & 256 & 0.5760 & 0.7102 & 0.6332 \\  \hline

\multirow{4}{*}{\makecell{CodeBERT-\\BiGRU}} & \multirow{1}{*}{$1e^{-4}$} & 128  &	0.3794 & 0.2923 & 0.5406 & 128 &	0.3794 & 0.2923 & 0.5406 & 128 &	0.3794 & 0.2923 & 0.5406\\ \cline{2-14} 

& \multirow{1}{*}{$1e^{-5}$} & 128  & 0.6032 & 0.6239 & 0.6201 & 128 & 0.6418 & 0.6680 & 0.6574 & 128 & 0.6014 & 0.6111 & 0.6120 \\ \cline{2-14} 

& \multirow{1}{*}{$1e^{-6}$} & 512  & 0.5741 & 0.6575 & 0.6201 & 512 &	0.5717 & 0.6489 & 0.6164 &  512 & 0.5828 & 0.6478 & 0.6212  \\ \cline{2-14}

& \multirow{1}{*}{$2e^{-5}$} & 512 & 0.6484 & 0.6582 & 0.6563  &  256 & 0.6615 & 0.6682 & 0.6673 & 256 & 0.6346 & 0.6597 & 0.6504 \\  \hline

\end{tabular}
\end{table*}

\begin{table*}[!t]
\caption{The best performance ($\mathbf{F1}_\psi$, $\mathbf{P}_\psi$, and $\mathbf{R}_\psi$) achieved by CodeT5-RNN models on defect detection dataset} \label{codet5_rnn_model_p_r_f1}
\centering
\begin{tabular}{p{1.2cm}|p{1.4cm}||p{.6cm}|p{.65cm}|p{.65cm}|p{.65cm} ||p{.6cm}|p{.65cm}|p{.65cm}|p{.65cm}||p{.6cm}|p{.65cm}|p{.65cm}|p{.65cm}}
\hline \hline

\multirow{2}{*}{\textbf{Model}} & \multirow{2}{*}{\textbf{\makecell{Learning \\Rate ($l$)}}} & \multicolumn{4}{c||} {\textbf{Adam}} & \multicolumn{4}{c||} {\textbf{NAdam}} & \multicolumn{4}{c}{\textbf{RMSprop}} \\ \cline{3-14}

 &  & RHUs & \centering $\mathbf{F1}_\psi$ & \centering $\mathbf{P}_\psi$ &  \makecell{$\mathbf{R}_\psi$}  & RHUs & \centering $\mathbf{F1}_\psi$ & \centering $\mathbf{P}_\psi$ & \makecell{$\mathbf{R}_\psi$}  & RHUs & \centering $\mathbf{F1}_\psi$ & \centering $\mathbf{P}_\psi$ &  \makecell{$\mathbf{R}_\psi$} \\ \hline \hline

\multirow{4}{*}{\makecell{CodeT5-\\LSTM}} & \multirow{1}{*}{$1e^{-4}$} &  512 &	0.6715 & 0.6714 & 0.6717  & 512 &	0.6763 & 0.6762 & 0.6764 & 256 &	0.6594	& 0.6635	& 0.6589\\ \cline{2-14} 

& \multirow{1}{*}{$1e^{-5}$} & 512 & 0.6543 & 0.6572 & 0.6537  & 512 & 0.6640 & 0.6650 & 0.6636 &   128 & 0.6671	& 0.6680	& 0.6691\\ \cline{2-14} 
& \multirow{1}{*}{$1e^{-6}$} & 128 & 0.5141 & 0.5642 & 0.5641 & 128 &	0.5289 & 0.5670 & 0.5681 & 256 & 0.5563	& 0.6125	& 0.5974\\ \cline{2-14}
  
& \multirow{1}{*}{$2e^{-5}$} & 512 & 0.6775 & 0.6781 & 0.6772 & 256 &   0.6685 & 0.6757 & 0.6742 & 256 & 0.6696	& 0.6714	& 0.6691\\ \hline

\multirow{4}{*}{\makecell{CodeT5-\\BiLSTM}} & \multirow{1}{*}{$1e^{-4}$} & 256 &	0.6667 & 0.6701 & 0.6662 & 128 &  0.6662 & 0.6720 & 0.6658 & 512 &	0.6576	& 0.6588 &	0.6600 \\  \cline{2-14} 

& \multirow{1}{*}{$1e^{-5}$} & 256  &	0.6595 & 0.6602 & 0.6614 & 256 & 0.6681 & 0.6753 & 0.6739 & 256 &	0.6614	& 0.6613	& 0.6622 \\ 
\cline{2-14} 

& \multirow{1}{*}{$1e^{-6}$} & 128 &	0.5182 & 0.5708 & 0.5681 & 512 & 0.5635 & 0.5915 & 0.5897 & 512 &	0.5487	& 0.6385	& 0.6032
 \\  \cline{2-14}
  
& \multirow{1}{*}{$2e^{-5}$} & 128 &	0.6669 & 0.6766 & 0.6739 & 256 & 	0.6618 & 0.6638 & 0.6647 &  	128 & 	0.6606	& 0.6650	& 0.6651 \\ \hline

\multirow{4}{*}{\makecell{CodeT5-\\GRU}} & \multirow{1}{*}{$1e^{-4}$} & 512  & 0.6718 & 0.6824 & 0.6790 & 512 & 0.6528 & 0.6630 & 0.6607  & 128 & 0.6681	& 0.6696	& 0.6706 \\ \cline{2-14} 

& \multirow{1}{*}{$1e^{-5}$} & 512  & 0.6737 & 0.6743 & 0.6753 & 256 & 0.6663 & 0.6678 & 0.6687 & 256 & 0.6617	& 0.6681	& 0.6673  \\  \cline{2-14} 

& \multirow{1}{*}{$1e^{-6}$} & 512  & 0.5145 & 0.5717 & 0.5677 & 256 & 0.5295 & 0.5625 & 0.5655 & 128 & 0.5459 & 	0.5791	& 0.5783\\  \cline{2-14}

& \multirow{1}{*}{$2e^{-5}$} & 128  & 	0.6665 & 0.6728 & 0.6662 & 512 & 0.6565 & 0.6586 & 0.6559  & 128 & 0.6647	& 0.6749	& 0.6720\\ \hline

\multirow{4}{*}{\makecell{CodeT5-\\BiGRU}} & \multirow{1}{*}{$1e^{-4}$} & 256  &  0.6650 & 0.6648 & 0.6654 & 512 &	0.6638 & 0.6656 & 0.6665 & 128 &	0.6673	& 0.6802	& 0.6757\\ \cline{2-14} 

& \multirow{1}{*}{$1e^{-5}$} & 512  & 0.6586 & 0.6588 & 0.6600 & 512 &	0.6554 & 0.6627 & 0.6552  & 256 & 0.6567	& 0.6567	& 0.6567 \\ \cline{2-14} 

& \multirow{1}{*}{$1e^{-6}$} & 256  & 0.5618 & 0.5615 & 0.5641 & 128 & 0.5150 & 0.5921 & 0.5761 & 128 & 0.5309	& 0.5934	& 0.5813 \\ \cline{2-14}

& \multirow{1}{*}{$2e^{-5}$} & 256 & 	0.6644 & 0.6730 & 0.66434  & 256 & 0.6631 & 0.6656 & 0.6625 & 256 & 0.6646	& 0.6645 & 0.6654 \\ 
\hline

\end{tabular}
\end{table*}

\begin{table*}[!t]
\caption{The best performance ($\mathbf{F1}_\psi$, $\mathbf{P}_\psi$, and $\mathbf{R}_\psi$) achieved by CodeT5$^+$-RNN models on defect detection dataset} \label{codet5+_rnn_model_p_r_f1}
\centering
\begin{tabular}{p{1.2cm}|p{1.4cm}||p{.6cm}|p{.65cm}|p{.65cm}|p{.65cm} ||p{.6cm}|p{.65cm}|p{.65cm}|p{.65cm}||p{.6cm}|p{.65cm}|p{.65cm}|p{.65cm}}
\hline \hline

\multirow{2}{*}{\textbf{Model}} & \multirow{2}{*}{\textbf{\makecell{Learning \\Rate ($l$)}}} & \multicolumn{4}{c||} {\textbf{Adam}} & \multicolumn{4}{c||} {\textbf{NAdam}} & \multicolumn{4}{c}{\textbf{RMSprop}} \\ \cline{3-14}

 &  & RHUs & \centering $\mathbf{F1}_\psi$ & \centering $\mathbf{P}_\psi$ &  \makecell{$\mathbf{R}_\psi$}  & RHUs & \centering $\mathbf{F1}_\psi$ & \centering $\mathbf{P}_\psi$ & \makecell{$\mathbf{R}_\psi$}  & RHUs & \centering $\mathbf{F1}_\psi$ & \centering $\mathbf{P}_\psi$ &  \makecell{$\mathbf{R}_\psi$} \\ \hline \hline

\multirow{4}{*}{\makecell{CodeT5$^+$-\\LSTM}} & \multirow{1}{*}{$1e^{-4}$} & 128 & 0.6289	& 0.6377	& 0.6373 & 256 &	0.6363	& 0.6463	& 0.6449 & 128 &	0.6451	& 0.6468	& 0.6446 \\  \cline{2-14}

& \multirow{1}{*}{$1e^{-5}$} & 128 & 0.6658	& 0.6746 &	0.6724 & 256 &	0.6659	& 0.6660	& 0.6658 & 512 &  0.6627	& 0.6640	& 0.6651\\ \cline{2-14}

& \multirow{1}{*}{$1e^{-6}$} & 128 & 0.6085	& 0.6368 & 	0.6285 & 256 &	0.6174 &	0.6173	& 0.6175 & 512 & 0.5976	& 0.6215	& 0.6168\\ \cline{2-14}
  
& \multirow{1}{*}{$2e^{-5}$} & 128 &	0.6617 &	0.6615 & 	0.6622 & 512 & 0.6603 & 0.6687	& 0.6686 & 256 & 0.6562	& 0.6632	& 0.6559\\ \hline

\multirow{4}{*}{\makecell{CodeT5$^+$-\\BiLSTM}} & \multirow{1}{*}{$1e^{-4}$} & 128  &	0.6331 &	0.6335	 & 0.6351 &	256 & 0.6447	& 0.6522 &	0.6515 & 512	& 0.6437	& 0.6458	& 0.6431\\  \cline{2-14} 

& \multirow{1}{*}{$1e^{-5}$} & 512 &	0.6718	& 0.6785	& 0.6772 & 512 &	0.6740	& 0.6773	& 0.6735 & 256 &	0.6741	& 0.6741	& 0.6750  \\ \cline{2-14} 

& \multirow{1}{*}{$1e^{-6}$} & 512 &	0.6176	& 0.6394 &	0.6336 & 512 &	0.6098	& 0.6356 &	0.6285 & 512 &	0.6056	& 0.6263 &	0.6223 \\ \cline{2-14}
  
& \multirow{1}{*}{$2e^{-5}$} & 128 &	0.6717 &	0.6716 &	0.6720 & 512 &	0.6505	& 0.6707	& 0.6526 & 256 &	0.6612	& 0.6635	& 0.6643 \\  \hline

\multirow{4}{*}{\makecell{CodeT5$^+$-\\GRU}} & \multirow{1}{*}{$1e^{-4}$} & 512  & 0.6198 &	0.6451 &	0.6237 & 512 & 0.6315	& 0.6326	& 0.6310 & 256 & 0.6380	& 0.6388	& 0.6376\\ \cline{2-14} 

& \multirow{1}{*}{$1e^{-5}$} & 512  & 0.6687	& 0.6695	& 0.6706 & 512 &	0.6661	& 0.6659	& 0.6665 & 128 & 0.6647	& 0.6648	& 0.6658 \\ \cline{2-14} 

& \multirow{1}{*}{$1e^{-6}$} & 512  & 0.6083	& 0.6378	& 0.6288 & 512 & 0.5989 &	0.6303 &	0.6215 & 512 & 0.5948	& 0.6171	& 0.6135 \\ \cline{2-14}

& \multirow{1}{*}{$2e^{-5}$} & 256 & 	0.6729	& 0.6789	& 0.6779 & 256 & 0.6677	& 0.6764	& 0.6676 & 256 & 0.6612	& 0.6658	& 0.6658 \\ 
\hline

\multirow{4}{*}{\makecell{CodeT5$^+$-\\BiGRU}} & \multirow{1}{*}{$1e^{-4}$} & 512  & 0.6432	& 0.6441	& 0.6428 & 128 & 0.6357	& 0.6405	& 0.6413 & 512 & 0.6549	& 0.6562	& 0.6574 \\ \cline{2-14} 

& \multirow{1}{*}{$1e^{-5}$} & 512  & 0.6661	& 0.6659	& 0.6665 & 256 	& 0.6652 &	0.6694	& 0.6647 &  512 & 0.6642	& 0.6644	& 0.6654 \\ \cline{2-14} 

& \multirow{1}{*}{$1e^{-6}$} & 256 & 0.6120	& 0.6345	& 0.6288 & 512 &	0.6226	& 0.6266	& 0.6281 & 256 & 0.6003	& 0.6210	& 0.6175  \\ \cline{2-14}

& \multirow{1}{*}{$2e^{-5}$} & 256 & 	0.6651	& 0.6769 &	0.6731  & 512 & 0.6684	& 0.6688	& 0.6698 & 256 & 0.6682	& 0.6843	& 0.6779 \\  \hline 

\end{tabular}
\end{table*}

\begin{figure} []
     \centering
     \begin{subfigure}[b]{0.49\linewidth}
         \captionsetup{justification=centering}
         \begin{tikzpicture}[scale=.95]
            \input{codebert_lstm_accuracy}
        \end{tikzpicture}
        \caption{CodeBERT-LSTM}
         \label{codebert_lstm_accuracy}
     \end{subfigure}
     \hfill
       \begin{subfigure}[b]{0.49\linewidth}
        \captionsetup{justification=centering}
         \begin{tikzpicture}[scale=.95]
            \input{codebert_bilstm_accuracy}
        \end{tikzpicture}
        \caption{CodeBERT-BiLSTM}
         \label{codebert_bilstm_accuracy}
     \end{subfigure}
     \hfill
     \begin{subfigure}[b]{0.49\linewidth}
        \captionsetup{justification=centering}
         \begin{tikzpicture}[scale=.95]
            \input{codebert_gru_accuracy}
        \end{tikzpicture}
       \caption{CodeBERT-GRU}
         \label{codebert_gru_accuracy}
     \end{subfigure}
     \hfill     
     \begin{subfigure}[b]{0.49\linewidth}
        \captionsetup{justification=centering}
         \begin{tikzpicture}[scale=.95]
            \input{codebert_bigru_accuracy}
        \end{tikzpicture}
       \caption{CodeBERT-BiGRU}
         \label{codebert_bigru_accuracy}
     \end{subfigure}
    \vspace{-5mm}


           \caption{$\mathbf{A_{avg}}$ scores of the CodeBERT-RNN models for code understanding on the defect detection dataset, evaluated with $l=\{1e^{-4}, ~2e^{-5}, ~1e^{-5}, ~1e^{-6}\}$, $\Delta=$\{AdamW, NAdam, RMSprop\}, and $h=\{128, 256, 512\}$.}
           
        \label{codebert_rnn_accuracy}
\end{figure}

\begin{figure} []
     \centering
     \begin{subfigure}[b]{0.49\linewidth}
         \captionsetup{justification=centering}
         \begin{tikzpicture}[scale=1]
            \input{codebert_lstm_f1_performance_with_lr}      
            \node[above,font=\tiny] at (current bounding box.north) {$\Delta=$ AdamW};
        \end{tikzpicture}
         \label{codebert_lstm_f1_performance_with_lr}
         \vspace{-5mm}
     \end{subfigure}
     \hfill
       \begin{subfigure}[b]{0.49\linewidth}
        \captionsetup{justification=centering}
         \begin{tikzpicture}[scale=1]
            \input{codebert_lstm_accuracy_performance_with_lr}
        \node[above,font=\tiny] at (current bounding box.north) {$\Delta=$ AdamW};
        \end{tikzpicture}
         \label{codebert_lstm_accuracy_performance_with_lr}
         \vspace{-5mm}
     \end{subfigure}
     \hfill
        \begin{subfigure}[b]{0.49\linewidth}
        \captionsetup{justification=centering}
         \begin{tikzpicture}[scale=1]
            \input{codebert_lstm_f1_performance_with_lr_nadam}
        \node[above,font=\tiny] at (current bounding box.north) {$\Delta=$ NAdam};
        \end{tikzpicture}
         \label{codebert_lstm_f1_performance_with_lr_nadam}
     \end{subfigure}
     \hfill
            \begin{subfigure}[b]{0.49\linewidth}
        \captionsetup{justification=centering}
         \begin{tikzpicture}[scale=1]
            \input{codebert_lstm_accu_performance_with_lr_nadam}
        \node[above,font=\tiny] at (current bounding box.north) {$\Delta=$ NAdam};
        \end{tikzpicture}
         \label{codebert_lstm_accu_performance_with_lr_nadam}
     \end{subfigure}
     \vspace{-10mm}


           \caption{Comparative analysis of $\mathbf{F1_{avg}}$ and $\mathbf{A_{avg}}$ for CodeBERT-RNN models using $\Delta=$\{AdamW, NAdam\}, $h=\{128, 256, 512\}$, and  $l=\{1e^{-4}, ~2e^{-5}, ~1e^{-5}, ~1e^{-6}\}$.}
        \label{codebert_rnn_f1_a}
\end{figure}

\subsection{Results}
    
Tables \ref{roberta_rnn_model_p_r_f1}-\ref{codet5+_rnn_model_p_r_f1} present the  $\mathbf{P}_\psi$,  $\mathbf{R}_\psi$, and $\mathbf{F1}_\psi$ scores that are achieved by the RoBERTa-, CodeBERT-, CodeT5-, and CodeT5$^+$-RNN models on the benchmark dataset. These experiments are conducted under specific hyperparameter settings including $l=\{1e^{-4}, ~2e^{-5}, ~1e^{-5}, ~1e^{-6}\}$, $epoch=5$, $h=\{128, ~256, ~512\}$, $d=0.1, 0.2$ and $\Delta$=\{\texttt{AdamW}, \texttt{NAdam}, \texttt{RMSprop}\}. Four notable RNN architectures—LSTM, GRU, BiGRU, and BiLSTM—are integrated with LLMs to evaluate model performance in code understanding. Table \ref{roberta_rnn_model_p_r_f1} presents the results of the four RoBERTa-RNN\footnote{RoBERTa-RNN refers to four models, each incorporating a different RNN variant: LSTM, GRU, BiLSTM, and BiGRU. The same applies to the CodeBERT-RNN, CodeT5-RNN, and CodeT5$^+$-RNN models.} models (RoBERTa-LSTM, RoBERTa-BiLSTM, RoBERTa-GRU, and RoBERTa-BiGRU), which demonstrate significant performance with learning rates of $l=1e^{-5}$ and $l=1e^{-6}$ in most cases. Furthermore, optimizers such as \texttt{AdamW} and \texttt{NAdam} play a crucial role in enhancing the models' performance. It is evident from  Table \ref{codebert_rnn_model_p_r_f1} that the CodeBERT-RNN models struggled to achieve competitive $\mathbf{P}_\psi$,  $\mathbf{R}_\psi$, and $\mathbf{F1}_\psi$ scores when using a learning rate of $l=1e^{-4}$. Thus, a lower $l$ negatively affects the performance of the CodeBERT-RNN models. In contrast, as shown in Tables \ref{codet5_rnn_model_p_r_f1} and \ref{codet5+_rnn_model_p_r_f1}, the CodeT5- and CodeT5$^+$-RNN models achieved outstanding $\mathbf{P}\psi$, $\mathbf{R}\psi$, and $\mathbf{F1}\psi$ scores. However, their performance was relatively lower when using a learning rate of $l=1e^{-4}$.


Figure \ref{roberta_rnn_accuracy} illustrates the average accuracy ($\mathbf{A_{avg}}$) of the RoBERTa-RNN models. The results show that the models achieved higher $\mathbf{A_{avg}}$ values when using learning rates of $l=1e^{-5}$ and $l=1e^{-6}$, combined with the optimizers $\Delta$=\texttt{Adam} and \texttt{NAdam}, in most cases. Among these, the RoBERTa-BiGRU model achieved the highest $\mathbf{A_{avg}}$ of 64.25\% with $l=1e^{-5}$ and $\Delta$=\texttt{NAdam}, as shown in Figure \ref{roberta_accuracy_LR_0.000001}. Moreover, Figure \ref{roberta_rnn_f1_a} provides a comparative analysis of the average $\mathbf{F1_{avg}}$ and $\mathbf{A_{avg}}$ scores for the RoBERTa-RNN models, highlighting that the best results are consistently obtained with a learning rate of $l=1e^{-5}$. Figure \ref{codebert_rnn_accuracy} demonstrates that the CodeBERT-RNN models obtained notable $\mathbf{A_{avg}}$ scores when using learning rates of $l=2e^{-5}$ and $l=1e^{-5}$, along with the optimizers $\Delta$=\texttt{Adam} and \texttt{NAdam}.  Among these, the CodeBERT-BiGRU model achieved the highest $\mathbf{A_{avg}}$ score of 64.91\% with $l=2e^{-5}$ and $\Delta$=\texttt{NAdam}, as shown in Figure \ref{codebert_bigru_accuracy}. Additionally, Figure \ref{codebert_rnn_f1_a} highlights that the CodeBERT-BiGRU model obtained the top $\mathbf{F1_{avg}}$  score using the same $l$ and $\Delta$ values.

\begin{figure} []
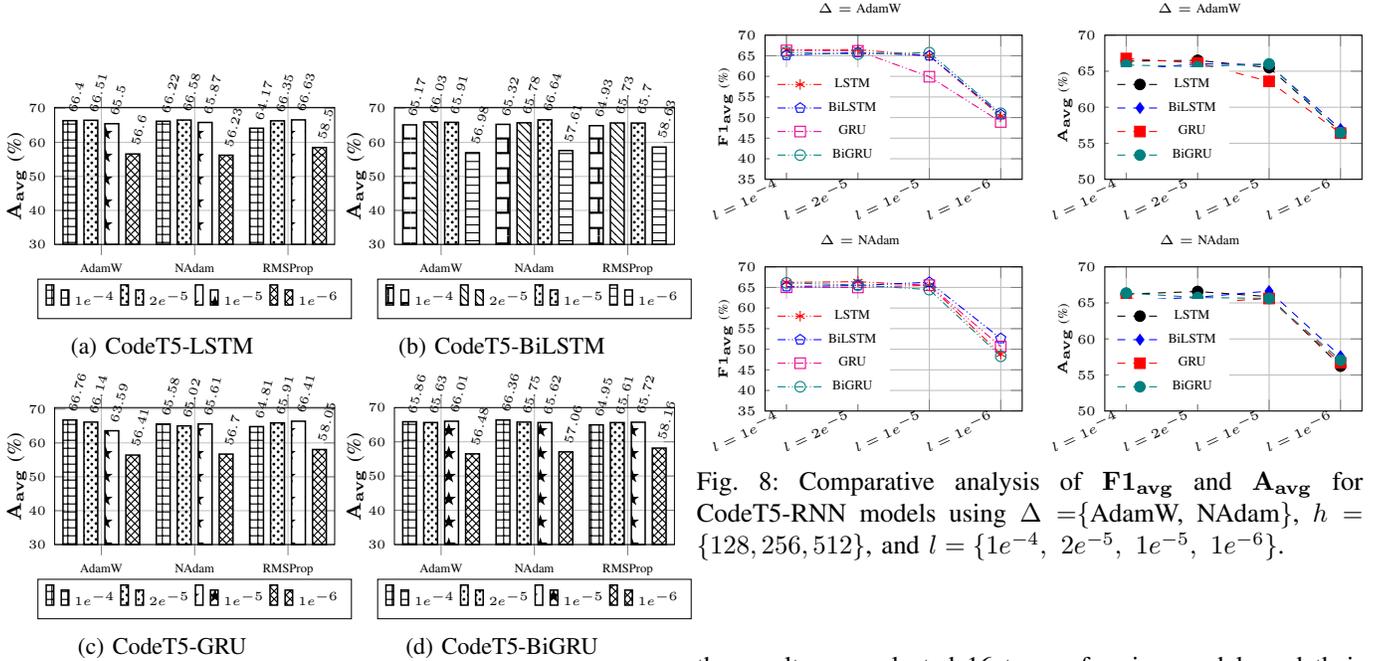

     \centering
     \begin{subfigure}[b]{0.49\linewidth}
         \captionsetup{justification=centering}
         \begin{tikzpicture}[scale=.95]
            \input{codet5_lstm_accuracy}
        \end{tikzpicture}
        \caption{CodeT5-LSTM}
         \label{codet5_lstm_accuracy}
     \end{subfigure}
     \hfill
       \begin{subfigure}[b]{0.49\linewidth}
        \captionsetup{justification=centering}
         \begin{tikzpicture}[scale=.95]
            \input{codet5_bilstm_accuracy}
        \end{tikzpicture}
        \caption{CodeT5-BiLSTM}
         \label{codet5_bilstm_accuracy}
     \end{subfigure}
     \hfill
     \begin{subfigure}[b]{0.49\linewidth}
        \captionsetup{justification=centering}
         \begin{tikzpicture}[scale=.95]
            \input{codet5_gru_accuracy}
        \end{tikzpicture}
       \caption{CodeT5-GRU}
         \label{codet5_gru_accuracy}
     \end{subfigure}
     \hfill     
     \begin{subfigure}[b]{0.49\linewidth}
        \captionsetup{justification=centering}
         \begin{tikzpicture}[scale=.95]
            \input{codet5_bigru_accuracy}
        \end{tikzpicture}
       \caption{CodeT5-BiGRU}
         \label{codet5_bigru_accuracy}
     \end{subfigure}
\vspace{-5mm}
    

  \caption{$\mathbf{A_{avg}}$ scores of the CodeT5-RNN models for code understanding on the defect detection dataset, evaluated with $l=\{1e^{-4}, ~2e^{-5}, ~1e^{-5}, ~1e^{-6}\}$, $\Delta=$\{AdamW, NAdam, RMSprop\}, and $h=\{128, 256, 512\}$.}
          
        \label{codet5_rnn_accuracy}
\end{figure}

\begin{figure} [!t]
     \centering
     \begin{subfigure}[b]{0.49\linewidth}
         \captionsetup{justification=centering}
         \begin{tikzpicture}[scale=1]
            \input{codet5_lstm_f1_performance_with_lr}      
            \node[above,font=\tiny] at (current bounding box.north) {$\Delta=$ AdamW};
        \end{tikzpicture}
         \label{codet5_lstm_f1_performance_with_lr}
         \vspace{-5mm}
     \end{subfigure}
     \hfill
       \begin{subfigure}[b]{0.49\linewidth}
        \captionsetup{justification=centering}
         \begin{tikzpicture}[scale=1]
            \input{codet5_lstm_accuracy_performance_with_lr}
        \node[above,font=\tiny] at (current bounding box.north) {$\Delta=$ AdamW};
        \end{tikzpicture}
         \label{codet5_lstm_accuracy_performance_with_lr}
         \vspace{-5mm}
     \end{subfigure}
     \hfill
        \begin{subfigure}[b]{0.49\linewidth}
        \captionsetup{justification=centering}
         \begin{tikzpicture}[scale=1]
            \input{codet5_lstm_f1_performance_with_lr_nadam}
        \node[above,font=\tiny] at (current bounding box.north) {$\Delta=$ NAdam};
        \end{tikzpicture}
         \label{codet5_lstm_f1_performance_with_lr_nadam}
     \end{subfigure}
     \hfill
            \begin{subfigure}[b]{0.49\linewidth}
        \captionsetup{justification=centering}
         \begin{tikzpicture}[scale=1]
            \input{codet5_lstm_accu_performance_with_lr_nadam}
        \node[above,font=\tiny] at (current bounding box.north) {$\Delta=$ NAdam};
        \end{tikzpicture}
         \label{codet5_lstm_accu_performance_with_lr_nadam}
     \end{subfigure}
     \vspace{-10mm}


           \caption{Comparative analysis of $\mathbf{F1_{avg}}$ and $\mathbf{A_{avg}}$ for CodeT5-RNN models using $\Delta=$\{AdamW, NAdam\}, $h=\{128, 256, 512\}$, and  $l=\{1e^{-4}, ~2e^{-5}, ~1e^{-5}, ~1e^{-6}\}$.}
        \label{codet5_rnn_f1_a}
\end{figure}

Figure \ref{codet5_rnn_accuracy} shows that the CodeT5-RNN models achieved competitive $\mathbf{A_{avg}}$ scores across all optimizers but performed lower when using a learning rate of $l=1e^{-6}$. This indicates that smaller $l$ values negatively impact model performance. Notably, the CodeT5 models performed better with the RNN variants LSTM and GRU. The CodeT5-GRU model achieved the highest $\mathbf{A_{avg}}$ score of 66.76\%, as shown in Figure \ref{codet5_gru_accuracy}. Additionally, Figure \ref{codet5_rnn_f1_a} provides a comparative analysis of the $\mathbf{F1_{avg}}$ and $\mathbf{A_{avg}}$ scores, highlighting that the models consistently achieved higher results except when using the learning rate $l=1e^{-6}$. Figure \ref{codet5plus_rnn_accuracy} demonstrates that the CodeT5$^+$ models achieved competitive results across all optimizers and learning rates. Among them, the CodeT5$^+$-BiLSTM model attained the highest $\mathbf{A_{avg}}$ score of 67.10\% with a learning rate of $l=1e^{-5}$ and $\Delta$=\texttt{RMSProp}, as shown in Figure \ref{codet5_plus_bilstm_accuracy}. Similarly, Figure \ref{codet5plus_rnn_f1_a} indicates that the models achieved higher $\mathbf{F1_{avg}}$ scores when using a learning rate of $l=1e^{-5}$.


\begin{figure} [!t]
     \centering
     \begin{subfigure}[b]{0.49\linewidth}
         \captionsetup{justification=centering}
         \begin{tikzpicture}[scale=.95]
            \input{codet5_plus_lstm_accuracy}
        \end{tikzpicture}
        \caption{CodeT5$^+$-LSTM}
         \label{codet5_plus_lstm_accuracy}
     \end{subfigure}
     \hfill
       \begin{subfigure}[b]{0.49\linewidth}
        \captionsetup{justification=centering}
         \begin{tikzpicture}[scale=.95]
            \input{codet5_plus_bilstm_accuracy}
        \end{tikzpicture}
        \caption{CodeT5$^+$-BiLSTM}
         \label{codet5_plus_bilstm_accuracy}
     \end{subfigure}
     \hfill
     \begin{subfigure}[b]{0.49\linewidth}
        \captionsetup{justification=centering}
         \begin{tikzpicture}[scale=.95]
            \input{codet5_plus_gru_accuracy}
        \end{tikzpicture}
       \caption{CodeT5$^+$-GRU}
         \label{codet5_plus_gru_accuracy}
     \end{subfigure}
     \hfill     
     \begin{subfigure}[b]{0.49\linewidth}
        \captionsetup{justification=centering}
         \begin{tikzpicture}[scale=.95]
            \input{codet5_plus_bigru_accuracy}
        \end{tikzpicture}
       \caption{CodeT5$^+$-BiGRU}
         \label{codet5_plus_bigru_accuracy}
     \end{subfigure}
    \vspace{-5mm}


  \caption{$\mathbf{A_{avg}}$ scores of the CodeT5$^+$-RNN models for code understanding on the defect detection dataset, evaluated with $l=\{1e^{-4}, ~2e^{-5}, ~1e^{-5}, ~1e^{-6}\}$, $\Delta=$\{AdamW, NAdam, RMSprop\}, and $h=\{128, 256, 512\}$.}
        \label{codet5plus_rnn_accuracy}
\end{figure}

\begin{figure} [!t]
     \centering
     \begin{subfigure}[b]{0.49\linewidth}
         \captionsetup{justification=centering}
         \begin{tikzpicture}[scale=1]
            \input{codet5_plus_lstm_f1_performance_with_lr}      
            \node[above,font=\tiny] at (current bounding box.north) {$\Delta=$ AdamW};
        \end{tikzpicture}
         \label{codet5_plus_lstm_f1_performance_with_lr}
         \vspace{-5mm}
     \end{subfigure}
     \hfill
       \begin{subfigure}[b]{0.49\linewidth}
        \captionsetup{justification=centering}
         \begin{tikzpicture}[scale=1]
            \input{codet5_plus_lstm_accuracy_performance_with_lr}
        \node[above,font=\tiny] at (current bounding box.north) {$\Delta=$ AdamW};
        \end{tikzpicture}
         \label{codet5_plus_lstm_accuracy_performance_with_lr}
         \vspace{-5mm}
     \end{subfigure}
     \hfill
        \begin{subfigure}[b]{0.49\linewidth}
        \captionsetup{justification=centering}
         \begin{tikzpicture}[scale=1]
            \input{codet5_plus_lstm_f1_performance_with_lr_nadam}
        \node[above,font=\tiny] at (current bounding box.north) {$\Delta=$ NAdam};
        \end{tikzpicture}
         \label{codet5_plus_lstm_f1_performance_with_lr_nadam}
     \end{subfigure}
     \hfill
            \begin{subfigure}[b]{0.49\linewidth}
        \captionsetup{justification=centering}
         \begin{tikzpicture}[scale=1]
            \input{codet5_plus_lstm_accu_performance_with_lr_nadam}
        \node[above,font=\tiny] at (current bounding box.north) {$\Delta=$ NAdam};
        \end{tikzpicture}
         \label{codet5_plus_lstm_accu_performance_with_lr_nadam}
     \end{subfigure}
    \vspace{-10mm}

           \caption{Comparative analysis of $\mathbf{F1_{avg}}$ and $\mathbf{A_{avg}}$ for CodeT5$^+$-RNN models using $\Delta=$\{AdamW, NAdam\}, $h=\{128, 256, 512\}$, and  $l=\{1e^{-4}, ~2e^{-5}, ~1e^{-5}, ~1e^{-6}\}$.}

        \label{codet5plus_rnn_f1_a}
\end{figure}

\begin{table*}[!t]
\caption{Comparison of accuracy and F1 scores between top-performing models (RoBERTa-RNN, CodeBERT-RNN, CodeT5-RNN, and CodeT5$^+$-RNN) and state-of-the-art models on the defect detection dataset.}
\centering
\begin{tabular}{c|c|c|c|c|c|c|c}
\hline 
 \multicolumn{2}{c|}{\textbf{Model}}  & \textbf{\makecell{Learning \\Rate ($l$)}} & \textbf{Optimizer ($\Delta$)} & \textbf{\makecell{Hidden \\Units ($h$)}} & \textbf{\textbf{A} (\%)} & \multicolumn{2}{c}{\textbf{F1 (\%)}} \\
\cline{1-2} \cline{7-8}
 \textbf{LLM} & \textbf{RNN} & & & & & \textbf{Weighted ($\psi$)} & \textbf{Macro ($\mu$)}\\
\hline\hline
- & BiLSTM \cite{zhou2019devign} & - & - & - & 59.37 & - & - \\ \hline
- & TextCNN \cite{zhou2019devign} & - & - & - & 60.69 & - & - \\ \hline

PLBART \cite{ahmad2021unified} & -  & - & - & - & 63.18  & - & - \\ \hline

RoBERTa \cite{zhou2019devign} & -  & - & - & - & 61.05 & - & - \\ \hline

RoBERTa + GFSA \cite{choi2023graph} & -  & - & - & - & 64.39 & - & - \\ \hline

CodeBERT \cite{zhou2019devign} & -  & - & - & - & 62.08 & - & - \\ \hline

CodeBERT + GFSA \cite{choi2023graph} & -  & - & - & - & 64.49 & - & - \\ \hline

CodeT5-Small \cite{wang2021codet5} & -  & - & - & - & 63.40 & - & - \\ \hline

CodeT5-Small + GFSA \cite{choi2023graph} & -  & - & - & - & 63.69 & - & - \\ \hline

C-BERT \cite{buratti2020exploring} & - & - & - & -& 65.45 & - & -\\ \hline

CoTexT \cite{phan2021cotext} & - & - & - & -& 66.62 & - & -\\ \hline

CodeT5-Base  & -  & - & - & - & 64.86 & 64.74 & - \\ \hline

CodeT5$^+$  & -  & - & - & - & 64.90 &	64.74 & - \\ \hline

\multirow{4}{*}{RoBERTa} & LSTM & {$1e^{-5}$} & AdamW & 128 &  64.46 & 63.72 & 63.0\\
\cline{2-8}
& BiLSTM & {$1e^{-5}$} & AdamW & 512 &  64.31 & 64.13 & 64.0\\
\cline{2-8}
& GRU & {$1e^{-5}$} & AdamW & 256 &  64.72 & 63.31 & 63.0\\
\cline{2-8}
& BiGRU & {$1e^{-5}$} & NAdam & 512 &  66.40 & 64.76 & 64.0\\
\hline

\multirow{4}{*}{CodeBERT} & LSTM & {$2e^{-5}$} & NAdam & 256 &  65.52 & 65.13 & 65.0\\
\cline{2-8}
& BiLSTM & {$2e^{-5}$} & NAdam & 256 &  65.70 & 65.71 & 65.0\\
\cline{2-8}
& GRU & {$2e^{-5}$} & AdamW & 512 &  66.03 & 65.32 & 65.0\\
\cline{2-8}
& BiGRU & {$2e^{-5}$} & NAdam & 512 &  65.63 & 65.20 & 65.0\\
\hline

\multirow{4}{*}{CodeT5} & LSTM & {$1e^{-5}$} & RMSProp & 128 &  66.91 & 66.71 & 66.0\\
\cline{2-8}
& BiLSTM & {$1e^{-5}$} & NAdam & 128 &  66.47 & 66.13 & 66.0\\
\cline{2-8}
& GRU & {$1e^{-4}$} & AdamW & 512 &  67.90 & 67.18 & 67.0\\
\cline{2-8}
& BiGRU & {$1e^{-4}$} & NAdam & 512 &  66.66 & 66.38 & 66.0\\
\hline

\multirow{4}{*}{CodeT5$^+$} & LSTM & {$2e^{-5}$} & NAdam & 512 &  66.91 & 66.87 & 67.0\\
\cline{2-8}
& BiLSTM & {$1e^{-5}$} & RMSProp & 256 &  67.50 & 67.41 & 67.0\\
\cline{2-8}
& GRU & {$2e^{-5}$} & NAdam & 256 &  66.76 & 66.77 & 67.0\\
\cline{2-8}
& BiGRU & {$2e^{-5}$} & RMSProp & 256 &  67.79 & 66.82 & 66.0\\
\hline
\end{tabular}
\label{top_models_on_defect_detection}

\vspace{1ex}
{ \scriptsize \textbf{Note:} We report the results from the referenced papers in terms of accuracy, as no other evaluation metrics and parameters were found for the defect detection dataset.}
\end{table*}

Table \ref{top_models_on_defect_detection} provides a comparative overview of the $\mathbf{A}$ and $\mathbf{F1}$ scores for the top-performing models and state-of-the-art benchmarks on the defect detection dataset. Based on the results, we selected 16 top-performing models and their corresponding parameters from RoBERTa-RNN, CodeBERT-RNN, CodeT5-RNN, and CodeT5$^+$-RNN. Additionally, the Table \ref{top_models_on_defect_detection} includes benchmark results from prior studies for models such as BiLSTM, TextCNN, PLBART, C-BERT, RoBERTa + GFSA, CodeBERT + GFSA, CodeT5-small, and CoTexT. These benchmarks serve as a reference, highlighting advancements in performance. The state-of-the-art models, including RoBERTa, CodeBERT, PLBART, and CodeT5-small, achieved notable $\mathbf{A}$ score of 61.05\%, 62.08\%, 63.18\%, and 63.40\%, respectively. Furthermore, their variants—such as CoTexT, C-BERT, RoBERTa+GFSA, and CodeT5-small+GFSA—demonstrated significant improvements. Within the reference models, CoTexT achieved the top accuracy, reaching 66.62\%. Meanwhile, the CodeT5-GRU model outperformed the reference models with an $\mathbf{A}$ of 67.90\% and an $\mathbf{F1}_\psi$ score of 67.18\%.

Furthermore, we conducted additional experiments using the top 16 models and their corresponding hyperparameter configurations (selected based on the results reported in Tables \ref{roberta_rnn_model_p_r_f1}–\ref{codet5+_rnn_model_p_r_f1}) across three real-world datasets: SearchAlg, SearchSortAlg, and SearchSortGTAlg. The weighted $\mathbf{P}\psi$, $\mathbf{R}\psi$, and $\mathbf{F1}_\psi$ scores from these experiments are summarized in Table \ref{roberta_codebert_codet5__rnn_model_p_r_f1_real_world_weighted}. The CodeT5-RNN models consistently outperformed other models in most cases, while CodeT5$^+$ also demonstrated competitive performance. The CodeT5-BiLSTM variant achieved the highest results, with an $\mathbf{F1}\psi$ score of 95.12\%, on the SearchAlg dataset. For the SearchSortAlg dataset, CodeT5-LSTM delivered the best performance, with $\mathbf{F1}\psi$ score of 96.72\%. Additionally, the RoBERTa-, CodeBERT-, and CodeT5$^+$-RNN models also performed competitively on the SearchAlg and SearchSortAlg datasets. In the SearchSortGTAlg dataset, the CodeT5$^+$-BiLSTM variant achieved the highest scores, with an $\mathbf{F1}_\psi$ score of 96.26\%. Table \ref{macro_roberta_codebert_codet5_rnn_model_p_r_f1_macro} shows the experimental results of $\mathbf{F1}\mu$, $\mathbf{P}\mu$, and $\mathbf{R}\mu$ scores across the same real-world datasets. The results clearly indicate that models based on CodeT5 and CodeT5$^+$ consistently achieved comparatively higher $\mathbf{F1}\mu$, $\mathbf{P}\mu$, and $\mathbf{R}\mu$ scores.

\begin{figure*}[]
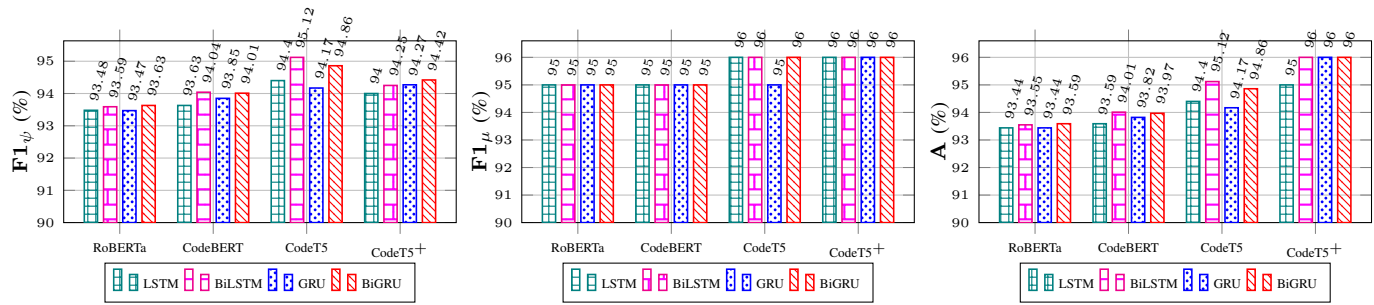

     \centering
     \begin{subfigure}[b]{0.32\linewidth}
         \captionsetup{justification=centering}
         \begin{tikzpicture}[scale=1]
            \input{F1_weighted_scores_SearchAlg}
        \end{tikzpicture}
         \label{F1_weighted_scores_SearchAlg}
     \end{subfigure}
     \hfill
       \begin{subfigure}[b]{0.32\linewidth}
        \captionsetup{justification=centering}
         \begin{tikzpicture}[scale=1]
            \input{F1_macro_scores_SearchAlg}
        \end{tikzpicture}
         \label{F1_macro_scores_SearchAlg}
     \end{subfigure}
     \hfill
     \begin{subfigure}[b]{0.32\linewidth}
        \captionsetup{justification=centering}
         \begin{tikzpicture}[scale=1]
            \input{accuracy_scores_SearchAlg}
        \end{tikzpicture}
         \label{accuracy_scores_SearchAlg}
     \end{subfigure}
     \vspace{-5mm}
          \caption{$\emph{\textbf{F1}}_\psi$, $\emph{\textbf{F1}}_\mu$, and $\emph{\textbf{A}}$ scores of the RoBERTa-RNN, CodeBERT-RNN, CodeT5-RNN, and CodeT5$^+$-RNN models on the SearchAlg dataset}
        \label{f1_a_searchalg}
\end{figure*}

\begin{figure*}[]
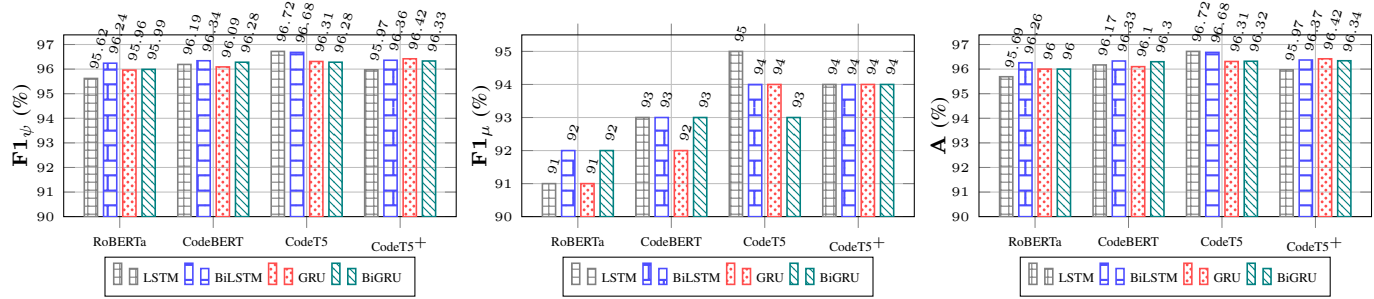

     \centering
     \begin{subfigure}[b]{0.32\linewidth}
         \captionsetup{justification=centering}
         \begin{tikzpicture}[scale=1]
            \input{F1_weighted_scores_SearchSortAlg}
        \end{tikzpicture}
         \label{F1_weighted_scores_SearchSortAlg}
     \end{subfigure}
     \hfill
       \begin{subfigure}[b]{0.32\linewidth}
        \captionsetup{justification=centering}
         \begin{tikzpicture}[scale=1]
            \input{F1_macro_scores_SearchSortAlg}
        \end{tikzpicture}
         \label{F1_macro_scores_SearchSortAlg}
     \end{subfigure}
     \hfill
     \begin{subfigure}[b]{0.32\linewidth}
        \captionsetup{justification=centering}
         \begin{tikzpicture}[scale=1]
            \input{accuracy_scores_SearchSortAlg}
        \end{tikzpicture}
         \label{accuracy_scores_SearchSortAlg}
     \end{subfigure}
     \vspace{-5mm}
          \caption{$\emph{\textbf{F1}}_\psi$, $\emph{\textbf{F1}}_\mu$, and $\emph{\textbf{A}}$ scores of the RoBERTa-RNN, CodeBERT-RNN, CodeT5-RNN, and CodeT5$^+$-RNN models on the SearchSortAlg dataset}
        \label{f1_a_searchsortalg}
\end{figure*}

Figures \ref{f1_a_searchalg}-\ref{f1_a_SearchSortGTAlg} depict a comprehensive performance comparison of the RoBERTa-, CodeBERT-, CodeT5-, and CodeT5$^+$-RNN models across the three real-world datasets: SearchAlg, SearchSortAlg, and SearchSortGTAlg. The evaluation metrics include the $\mathbf{F1}\psi$, $\mathbf{F1}\mu$, and $\mathbf{A}$. In Figure \ref{f1_a_searchalg}, CodeT5-BiLSTM and CodeT5-BiGRU achieved impressive $\mathbf{F1}\psi$ scores of 95.12\% and 94.86\%, respectively. The CodeT5- and CodeT5$^+$-RNN models demonstrated higher $\mathbf{F1}\mu$ scores compared to the RoBERTa- and CodeBERT-based models. A similar trend was observed in the accuracy comparison, where CodeT5- and CodeT5$^+$-RNN models excelled. Notably, the CodeT5$^+$-BiLSTM, GRU, and BiGRU variants achieved approximately 96\% accuracy, outperforming other models. In Figure \ref{f1_a_searchsortalg}, most models achieved competitive $\mathbf{F1}\psi$ and $\mathbf{A}$ scores. However, the CodeT5-LSTM model stood out with an $\mathbf{F1}\mu$ score of 95\%, surpassing the others. In Figure \ref{f1_a_SearchSortGTAlg}, despite the larger size, greater number of classes, and higher diversity of the SearchSortGTAlg dataset compared to the others, most LLM-RNN models still delivered high $\mathbf{F1}_\psi$ and $\mathbf{A}$ scores.

\begin{table*}[!t]
\caption{Quantitative results of $\mathbf{F1}\psi$, $\mathbf{P}\psi$, and $\mathbf{R}_\psi$  with the top LLM-RNN models on real-world datasets. }
\label{roberta_codebert_codet5__rnn_model_p_r_f1_real_world_weighted}
\centering
\begin{tabular}{p{1.3cm}|p{1.2cm}||p{1.3cm}|p{.9cm}|p{.9cm}||p{.9cm} |p{.9cm}|p{.9cm}||p{.9cm}|p{.9cm}|p{.9cm}}
\hline \hline

\multirow{2}{*}{\textbf{\makecell{LLM}}} & \multirow{2}{*}{\textbf{\makecell{RNN}}} & \multicolumn{3}{c||}{SearchAlg Dataset}  & \multicolumn{3}{c||}{SearchSortAlg Dataset} & \multicolumn{3}{c}{SearchSortGTAlg Dataset} \\ \cline{3-11}
 &  & \centering $\mathbf{F1}_\psi$ & \centering $\mathbf{P}_\psi$ &  \makecell{$\mathbf{R}_\psi$}  & \centering $\mathbf{F1}_\psi$ & \centering $\mathbf{P}_\psi$ &  \makecell{$\mathbf{R}_\psi$}  & \centering $\mathbf{F1}_\psi$ & \centering $\mathbf{P}_\psi$ &  \makecell{$\mathbf{R}_\psi$} \\ \hline \hline

\multirow{4}{*}{\makecell{RoBERTa}} & {LSTM}  &  0.934777 &	0.936710 &	0.934357  &	 0.956162 &	0.957187 &	0.956904 &	 0.951937 &	0.955060 &	  0.951875\\ \cline{2-11} 

& {Bi-LSTM}  & 0.935868 &	0.937218 &	0.935509  &	 0.962446 &	0.962522 &	0.962601  &  0.953376 &	0.954600 &	  0.953632  \\ \cline{2-11} 

& {GRU}  & 0.934721 &	0.935933 &	0.934357 &	0.959605 &	0.960074 &	0.960000  &  0.954208 &	0.955380 &	  0.954469  \\ \cline{2-11}

& {Bi-GRU}  & 0.936331 &	0.939025 &	0.935893 & 0.959934 &	0.960231 &	0.960000 &  0.959987 &	0.960998 &	  0.959993\\ \hline

\multirow{4}{*}{\makecell{CodeBERT}} & {LSTM}  &  0.936326 &	0.937258 &	0.935893  &	0.961870 &	0.962218 &	0.961734 & 0.960361 &	0.962104 &    0.960244 \\ \cline{2-11} 

& {Bi-LSTM}  & 0.940408 	& 0.941101 &	0.940115  &	0.963381 &	0.963735 &	0.963344 &  0.957453 &	0.958576 &	0.957482 \\ \cline{2-11} 

& {GRU}  & 0.938539 &	0.939108 &	0.938196 &	 0.960851 &	0.961069 &	0.960991  & 0.960394 &	0.961325 & 	0.960495  \\ \cline{2-11}

& {Bi-GRU}  & 0.940050 &	0.941520 &	0.939731 & 0.962837 &	0.962933 &	0.962972 & 0.957069 &	0.958253 &	0.957482 \\ \hline

\multirow{4}{*}{\makecell{CodeT5}} & {LSTM}  &  0.944000 &  0.944100 & 0.944000  &	0.967200 & 0.967600 & 0.967200 & 0.959800 & 0.961100 & 0.960700 \\ \cline{2-11} 

& {Bi-LSTM}  & 0.951200 & 0.951500 & 0.951200  & 0.966800 & 0.966800 & 0.966800  &  0.960100 & 0.960600 & 0.960700 \\ \cline{2-11} 

& {GRU}  & 0.941700 & 0.942500 & 0.941700 &	0.963100 & 0.964300 & 0.963100  &  0.952500 & 0.952600 & 0.954400  \\ \cline{2-11}

& {Bi-GRU}  & 0.948600 & 0.948600 & 0.948600 & 0.962800 & 0.962900 & 0.963200   &  0.949800 & 0.951500 & 0.950600 \\ \hline

\multirow{4}{*}{\makecell{CodeT5$^+$}} & {LSTM}  &  0.940024	& 0.940155 &	0.939923  &	0.959684 &	0.960487 &	0.959690  &	0.960484	& 0.961686 &	0.960747 \\ \cline{2-11} 

& {Bi-LSTM}  & 0.942474	& 0.942821 &	0.942418  &	0.963580	& 0.963708 &	0.963715 &  0.962622	& 0.963079	& 0.962713 \\ \cline{2-11} 

& {GRU}  & 0.942708	& 0.942975	& 0.942610 &	  0.964187 &	0.964375	& 0.964211 &  0.959239	& 0.960571	& 0.959156   \\ \cline{2-11}

& {Bi-GRU}  & 0.944225	& 0.944322	& 0.944146 &  0.963328	& 0.963863 &	0.963406   & 0.960338	& 0.961532	& 0.960328 \\ \hline

\end{tabular}
\end{table*}

\begin{table*}[]
\caption{Quantitative results of $\mathbf{F1}_\mu$, $\mathbf{P}_\mu$, and $\mathbf{R}_\mu$ with top LLM-RNN models on real-world datasets.}

\label{macro_roberta_codebert_codet5_rnn_model_p_r_f1_macro}
\centering
\begin{tabular}{p{1.3cm}|p{1.2cm}||p{1.3cm}|p{.9cm}|p{.9cm}||p{.9cm} |p{.9cm}|p{.9cm}||p{.9cm}|p{.9cm}|p{.9cm}}
\hline \hline

\multirow{2}{*}{\textbf{\makecell{LLM}}} & \multirow{2}{*}{\textbf{\makecell{RNN}}} & \multicolumn{3}{c||}{SearchAlg Dataset}  & \multicolumn{3}{c||}{SearchSortAlg Dataset} & \multicolumn{3}{c}{SearchSortGTAlg Dataset} \\ \cline{3-11}
 &  & \centering $\mathbf{F1}_\mu$ & \centering $\mathbf{P}_\mu$ &  \makecell{$\mathbf{R}_\mu$}  & \centering $\mathbf{F1}_\mu$ & \centering $\mathbf{P}_\mu$ &  \makecell{$\mathbf{R}_\mu$}  & \centering $\mathbf{F1}_\mu$ & \centering $\mathbf{P}_\mu$ &  \makecell{$\mathbf{R}_\mu$} \\ \hline \hline

\multirow{4}{*}{\makecell{RoBERTa}} & {LSTM}  &  0.95  & 0.94    &  0.95  &	0.91  &  0.91 &     0.94 &	 0.89  & 0.89    &  0.92 \\ \cline{2-11} 

& {Bi-LSTM}  & 0.95   &   0.95  &    0.95  &	0.92 & 0.92 &     0.93 &   0.90 &  0.90    &  0.91  \\ \cline{2-11} 

& {GRU}  & 0.95   &   0.95  &    0.95 &	 0.91 & 0.91  &     0.92 &   0.90 & 0.89   &   0.92  \\ \cline{2-11}

& {Bi-GRU}  & 0.95   &   0.95  &    0.95 & 0.92    &  0.92 &     0.92 &  0.92 & 0.91   &   0.93 \\ \hline

\multirow{4}{*}{\makecell{CodeBERT}} & {LSTM}  &  0.95   &   0.95    &  0.95  &	0.93 & 0.93  &    0.92 & 0.92  & 0.91  &  0.93 \\ \cline{2-11} 

& {Bi-LSTM}  & 0.95   &   0.95    &  0.95  &	0.93 & 0.92 &     0.93 &  0.91 & 0.91   &   0.93 \\ \cline{2-11} 

& {GRU}  & 0.95  &   0.95  &    0.96 &	0.92 &  0.92 &     0.93 &  0.92 &  0.92   &   0.93  \\ \cline{2-11}

& {Bi-GRU}  & 0.95   &   0.95    &  0.95 & 0.93 & 0.93   &   0.94 &  0.91 & 0.91   &   0.92 \\ \hline

\multirow{4}{*}{\makecell{CodeT5}} & {LSTM}  &  0.96   &   0.96  &    0.96  &	0.95 &     0.95     & 0.95 &	0.91 &  0.93   &   0.91  \\ \cline{2-11} 

& {Bi-LSTM}  & 0.96   &   0.96   &   0.96  &	0.94   &   0.94    &  0.94 &  0.91 & 0.92   &   0.91 \\ \cline{2-11} 

& {GRU}  & 0.95 & 0.95   &   0.95 &	  0.94 &  0.95 &     0.94 &  0.88 & 0.89   &   0.88  \\ \cline{2-11}

& {Bi-GRU}  & 0.96   &   0.96   &   0.96 &  0.93 &  0.94  & 0.92   & 0.89  & 0.90   &   0.89 \\ \hline

\multirow{4}{*}{\makecell{CodeT5$^+$}} & {LSTM}  &  0.96 &  0.96   &   0.95  &	0.94 &     0.94  &    0.94 &	 0.91  & 0.93   &   0.91 \\ \cline{2-11} 

& {Bi-LSTM}  & 0.96    &  0.96  &    0.96  &  0.94    &  0.94 &     0.94  &   0.92   & 0.93   &   0.92 \\ \cline{2-11} 

& {GRU}  & 0.96    &  0.96 &     0.96 &	 0.94    &  0.94 &     0.94   &  0.92 & 0.93 & 0.92   \\ \cline{2-11}

& {Bi-GRU}  & 0.96   &   0.96 &     0.96 & 0.94     & 0.94     & 0.94    &  0.92 & 0.93 & 0.92 \\ \hline

\end{tabular}
\end{table*}

\begin{figure*}[]
     \centering
     \begin{subfigure}[b]{0.32\linewidth}
         \captionsetup{justification=centering}
         \begin{tikzpicture}[scale=1]
            \input{F1_weighted_scores_SearchSortGTAlg}
        \end{tikzpicture}
         \label{F1_weighted_scores_SearchSortGTAlg}
     \end{subfigure}
     \hfill
       \begin{subfigure}[b]{0.32\linewidth}
        \captionsetup{justification=centering}
         \begin{tikzpicture}[scale=1]
            \input{F1_macro_scores_SearchSortGTAlg}
        \end{tikzpicture}
         \label{w_bilstm_1_comparison}
     \end{subfigure}
     \hfill
     \begin{subfigure}[b]{0.32\linewidth}
        \captionsetup{justification=centering}
         \begin{tikzpicture}[scale=1]
            \input{accuracy_scores_SearchSortGTAlg}
        \end{tikzpicture}
         \label{a_bilstm_1_comparison}
     \end{subfigure}
     \vspace{-5mm}
          \caption{$\emph{\textbf{F1}}_\psi$, $\emph{\textbf{F1}}_\mu$, and $\emph{\textbf{A}}$ scores of the RoBERTa-RNN, CodeBERT-RNN, CodeT5-RNN, and CodeT5$^+$-RNN models on the SearchSortGTAlg dataset}
        \label{f1_a_SearchSortGTAlg}
\end{figure*}

\section{Discussion} \label{final_discussion}


In this section, we analyze the performance of hybrid models and address the research questions. We also investigate how hyperparameters affect model performance. Lastly, we discuss the adaptability of the LLM-RNN approach and outline research limitations.

\subsection{Performance Analysis}

The objective of this study is to investigate how RNN integration further reinforces the contextual embeddings of LLM for code understanding tasks. 
To achieve this, comprehensive experiments are conducted using a variety of hyperparameter configurations on both benchmark and real-world datasets. Tables \ref{roberta_rnn_model_p_r_f1} through \ref{codet5+_rnn_model_p_r_f1} present the experimental results, showcasing model performance on the defect detection dataset. The experiments utilized learning rates ($l = 1e^{-4}, 2e^{-5}, 1e^{-5}, 1e^{-6}$), optimizers ($\Delta =$ \texttt{AdamW}, \texttt{NAdam}, \texttt{RMSprop}), and RHUs ($h = 128,~ 256,~ 512$). These results provide valuable insights into the models' behavior under various conditions, enabling us to identify and select the optimal configurations for code comprehension tasks.

Figures \ref{roberta_rnn_accuracy} and \ref{roberta_rnn_f1_a} illustrate $\mathbf{A_{avg}}$ and $\mathbf{F1_{avg}}$ of the RoBERTa-RNN models under various hyperparameter configurations. The results clearly demonstrate that reinforcing embeddings with RNN enhances RoBERTa's performance on the defect detection dataset. Notably, the RoBERTa-BiGRU model achieved an $\mathbf{A_{avg}}$ of 64.25\% (Figure \ref{roberta_accuracy_LR_0.000001}), reflecting a significant improvement of approximately 3.20\% ($\uparrow$) compared to the stand-alone RoBERTa model \cite{zhou2019devign}. Similarly, Figures \ref{codebert_rnn_accuracy} and \ref{codebert_rnn_f1_a} highlight the performance of the CodeBERT-RNN models on the same dataset. 
Among the evaluated models, the top-performing CodeBERT-BiGRU achieved an $\mathbf{A_{avg}}$ score of 64.91\%, representing an improvement of approximately 0.66\% ($\uparrow$) over the RoBERTa-BiGRU model and 3.86\% ($\uparrow$) over the stand-alone RoBERTa model. Moreover, CodeBERT-BiGRU demonstrates a 2.83\% ($\uparrow$) improvement in accuracy when compared to its base model \cite{zhou2019devign}. Also, the CodeBERT-BiGRU model achieved the highest $\mathbf{F1_{avg}}$ score of 64.45\%, outperforming other CodeBERT-RNN variants. 

CodeT5, an encoder-decoder model based on the T5 architecture \cite{raffel2020exploring}, is trained on CodeSearchNet and extended datasets. It is highly versatile, excelling in both code understanding and generation tasks, making it a specialized model for these applications. Figures \ref{codet5_rnn_accuracy} and \ref{codet5_rnn_f1_a} present the $\mathbf{A_{avg}}$ and $\mathbf{F1_{avg}}$ scores of the CodeT5-RNN models across different configurations. Our experiments reveal that CodeT5, combined with various RNN architectures, achieved significant $\mathbf{A_{avg}}$ scores, with most models reaching an average of 66.50($\pm$0.60)\%. In particular, the CodeT5-GRU model achieved an $\mathbf{A_{avg}}$ score of 66.76\% (Figure \ref{codet5_gru_accuracy}), marking improvements of approximately 3.36\% ($\uparrow$) and 1.90\% ($\uparrow$) compared to the CodeT5-small \cite{wang2021codet5} and CodeT5-base models, respectively (see in Table \ref{top_models_on_defect_detection}). This score also outperforms the top-performing CodeBERT-BiGRU model by approximately 1.85\% ($\uparrow$), underscoring the effectiveness of CodeT5 in code understanding tasks. Moreover, the CodeT5-LSTM model achieved the highest $\mathbf{F1_{avg}}$ score of 66.49\%, outperforming other variants. 

We also experimented with an advanced version of CodeT5, known as CodeT5$^+$, for our code understanding task. CodeT5$^+$ incorporates advanced transformer architectures, such as UL2 \cite{tay2022ul2}, and is trained on larger and more diverse datasets, enhancing its capabilities for code-related tasks. 
Figures \ref{codet5plus_rnn_accuracy} and \ref{codet5_rnn_f1_a} demonstrate that reinforcing CodeT5$^+$ embeddings with sequential RNN reprocessing leads to additional performance improvements. Specifically, the CodeT5$^+$-BiGRU model achieved an $\mathbf{A_{avg}}$ score of 66.80\% (Figure \ref{codet5_plus_bigru_accuracy}), reflecting approximately  1.90\% ($\uparrow$) improvement over the original CodeT5$^+$-base model. Table \ref{top_models_on_defect_detection} provides comparative results, demonstrating that our top-performing models significantly outperform existing state-of-the-art models by a notable margin.

\begin{mdframed}
\textbf{RQ1:} \emph{Does reinforcing contextual embeddings with RNN improve model performance?} \\

\noindent\textbf{Response:} Yes. Extensive experimental results on both benchmark and real-world datasets demonstrate that reprocessing LLM-generated contextual embeddings with sequential RNN architectures consistently improves code understanding performance. 
Across multiple LLM backbones (RoBERTa, CodeBERT, CodeT5, and CodeT5$^+$), the hybrid LLM–RNN models achieve superior accuracy and F1-scores compared to their standalone counterparts, confirming the effectiveness of the proposed embedding reinforcement strategy.
\end{mdframed}

To further evaluate the effectiveness of the models, the sixteen (16) top-performing LLM-RNN models are selected 
for experiments on three real-world datasets, as detailed in Tables \ref{roberta_codebert_codet5__rnn_model_p_r_f1_real_world_weighted} and \ref{macro_roberta_codebert_codet5_rnn_model_p_r_f1_macro}. 
Figures \ref{f1_a_searchalg}, \ref{f1_a_searchsortalg}, and \ref{f1_a_SearchSortGTAlg} provide a comparative analysis of $\mathbf{F1}\psi$, $\mathbf{F1}\mu$, and $\mathbf{A}$ scores across the three datasets. For the SearchAlg dataset, the CodeT5-RNN model achieved $\mathbf{F1}\psi$, $\mathbf{F1}\mu$, and $\mathbf{A}$ scores of 95.12\%, 96.00\%, and 95.12\%, respectively, outperforming all other models. A similar trend was observed for the SearchSortAlg and SearchSortGTAlg datasets, where CodeT5-RNN models consistently delivered superior performance. These results clearly demonstrate that integrating RNN architectures with CodeT5 significantly enhances its performance in code comprehension tasks.

\begin{mdframed}
\textbf{RQ2:}  \emph{How do hybrid LLM-RNN models perform on benchmark and real-world datasets in terms of code comprehension and debugging efficiency?} \\

\noindent\textbf{Response:} To address this \textbf{RQ2}, we utilized benchmark and real-world datasets for code comprehension tasks like defect detection and algorithm identification. The integration of RNNs with LLMs significantly improved defect detection performance compared to base models. For algorithm identification, the proposed models demonstrated notable performance improvements across the datasets.
\end{mdframed}

\begin{figure}[h]
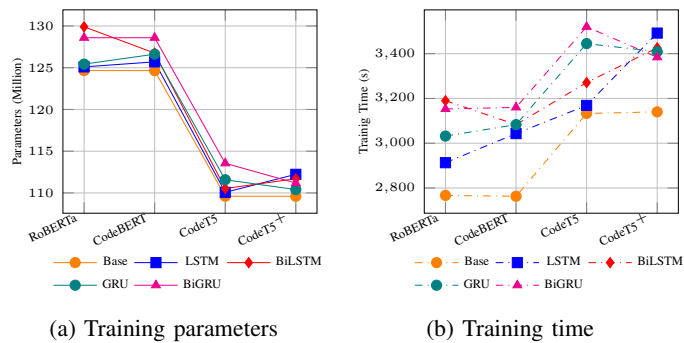

     \centering
     \begin{subfigure}[b]{0.48\linewidth}
     \centering
         \captionsetup{justification=centering}
         \begin{tikzpicture}[scale=.96]
            \input{training_parameters}
        \end{tikzpicture}
        \caption{Training parameters}
         \label{training_parameters}
     \end{subfigure}
     \hfill
       \begin{subfigure}[b]{0.48\linewidth}
       \centering
        \captionsetup{justification=centering}
         \begin{tikzpicture}[scale=.96]
            \input{time_curve}
        \end{tikzpicture}
        \caption{Training time}
         \label{time_curve}
     \end{subfigure}

          \caption{Comparison of training parameters and time between the top-performing models and their respective base models.}
        \label{training_parameters_times}
\end{figure}

Figure \ref{training_parameters_times} illustrates the training parameters and training times for the top-performing models and their base models. Notably, LLMs with BiGRU have the highest number of training parameters. Among them, the RoBERTa and CodeBERT models each have approximately 125 million parameters, while the CodeT5 and CodeT5$^+$ models contain around 112 million parameters, as shown in Figure \ref{training_parameters}. Despite having fewer training parameters, the CodeT5 and CodeT5$^+$ models, when combined with all RNN variants, required significantly more training time compared to the RoBERTa and CodeBERT models, as depicted in Figure \ref{time_curve}.

\subsection{Impact of Parameter Tuning on Model Performance}

We conducted extensive experiments involving hyperparameter tuning across different datasets to identify suitable models and parameters for our task. A total of approximately 642 experiments are performed, including 576 experiments on the defect detection dataset [$4 (l) \times 3 (h) \times 3 (\Delta) \times 4 (\text{RNNs}) \times 4 (\text{LLMs})$] and 48 experiments on real-world datasets [$4 (\text{LLMs}) \times 4 (\text{RNNs}) \times 3 (\text{Datasets})$]. Tables \ref{roberta_rnn_model_p_r_f1}-\ref{codet5+_rnn_model_p_r_f1} present the comprehensive results of fine-tuning the parameters on the benchmark dataset. Hyperparameter tuning significantly impacted model performance. For instance, as shown in Figure \ref{roberta_accuracy_LR_0.0001}, the RoBERTa-LSTM model achieved the highest $\mathbf{A_{avg}}$ of 63.60\% when using the \texttt{NAdam} optimizer and a learning rate of $l=1e^{-5}$. In contrast, the same model failed to deliver optimal results with other hyperparameter combinations. Similarly, Figure \ref{roberta_accuracy_LR_0.000001} demonstrates that the RoBERTa-BiGRU model achieved its highest $\mathbf{A_{avg}}$ score of 64.25\% with the \texttt{NAdam} optimizer and a learning rate of $l=1e^{-5}$, while its second-best result of 63.76\% was obtained using the \texttt{Adam} optimizer with the same learning rate. These findings highlight the importance of hyperparameter tuning in identifying optimal models. Without parameter fine-tuning, achieving competitive performance is challenging, as evidenced by the RoBERTa-BiGRU model's inability to perform well with other hyperparameter configurations.

Figure \ref{codet5_bigru_accuracy} shows that the CodeT5-GRU model achieved its lowest $\mathbf{A_{avg}}$ score of 56.41\% when using the \texttt{Adam} optimizer with a learning rate of $l=1e^{-6}$. This result is approximately 10.35\% ($\downarrow$) lower than the model's best performance, achieved with the same optimizer but a higher learning rate of $l=1e^{-4}$. These findings underscore that adjusting the learning rate ($l$) can significantly impact the model's performance. 
Table \ref{top_models_on_defect_detection} presents the top-performing 16 models (4 LLMs × 4 RNNs) along with their corresponding hyperparameters on the defect detection dataset. For instance, it can be seen that the CodeT5-LSTM/BiLSTM models achieved the best performance with a lower number of RHUs ($h=124$), whereas the CodeT5-GRU/BiGRU models performed better with a higher number of RHUs ($h=512$). This highlights the significant impact of the number of RHUs ($h$) on model performance. These top-performing models and their hyperparameters are subsequently used for experiments on real-world datasets, as detailed in Tables \ref{roberta_codebert_codet5__rnn_model_p_r_f1_real_world_weighted} and \ref{macro_roberta_codebert_codet5_rnn_model_p_r_f1_macro}.

\subsection{Adaptability}


Pretrained LLMs have achieved impressive performance across a wide range of language modeling tasks. Nevertheless, in domain-specific scenarios—particularly those involving strongly order-dependent data—general-purpose LLMs often struggle to reach state-of-the-art results. A key limitation lies in their contextual embeddings, which lack a sufficiently strong positional inductive bias, especially for long and highly structured sequences \cite{wu2025emergence}. This weakness can contribute to the well-known \emph{lost in the middle} phenomenon. In this paper, we demonstrate that reinforcing LLM contextual embeddings with RNNs significantly improves code comprehension performance compared to standalone base models. For example,  the CodeBERT model achieved an $\mathbf{A}$ of 62.08\%, while the CodeBERT-GRU model achieved 66.03\%, reflecting a 3.95\% ($\uparrow$) improvement. These results suggest that the proposed approach can be adapted to various application domains, including biomedical, healthcare, education, business, sentiment analysis, and entertainment.

\begin{mdframed}
\textbf{RQ3:} \emph{What are the advantages of hybrid models over stand-alone LLMs in handling long-term dependencies and contextual relationships within complex code structures?} \\

\noindent\textbf{Response:} While LLMs generate rich contextual embeddings through self-attention mechanisms, their positional inductive bias may limit fine-grained modeling of sequential and order-sensitive dependencies in highly structured source code. 
The hybrid models introduce an explicit sequential inductive bias that reinforces structural and long-range dependencies through recurrent processing. This complementary modeling enables a more refined representation of contextual relationships within complex code structures. Experimental results consistently demonstrate that the hybrid LLM–RNN models outperform their standalone LLM counterparts across benchmark and real-world datasets, confirming the effectiveness of sequential reinforcement in enhancing code comprehension performance.
\end{mdframed}

\subsection{Limitations}

We evaluate the performance of various hybrid models, combining LLMs and RNNs, for code understanding tasks using benchmark and real-world datasets. These models achieved notable improvements in tasks such as defect detection and algorithm identification compared to standalone LLMs. However, while the results are significant, they may vary due to the following reasons: ($i$) hyperparameter settings and their specific values, ($ii$) inconsistencies in the datasets, ($iii$) implementation factors, ($iv$) different combinations of LLMs and RNNs, and ($v$) differences in the overall model architecture.

\section{Conclusion} \label{conclusion_study}

This paper presented a systematic investigation of hybrid LLM–RNN architectures for enhanced code comprehension. By reprocessing LLM-generated contextual embeddings through sequential RNN, the proposed framework reintroduces explicit sequential inductive bias, enabling improved modeling of structural, order-sensitive, and long-range dependencies inherent in source code. 
Empirical results across both benchmark and large-scale real-world datasets consistently demonstrate that reinforcing contextual embeddings with RNN architectures yields significant and statistically consistent performance gains over standalone LLMs. On the defect detection benchmark, RoBERTa-BiGRU improved accuracy by approximately 5.35\%, CodeBERT-GRU by 3.95\%, and CodeT5$^+$-GRU by 4.50\% relative to their respective base models. The CodeT5-GRU model achieved the highest performance with $\mathbf{F1}_\psi = 67.18\%$, $\mathbf{F1}_\mu = 67.00\%$, and $\mathbf{A} = 67.90\%$. Furthermore, CodeT5-RNN models consistently outperformed other configurations on three real-world datasets, achieving weighted F1-scores exceeding 95\% in algorithm identification tasks, which demonstrates strong robustness and generalization capability. 
Given these promising outcomes, this approach could be adapted to address more complex challenges in domains such as medical diagnostics, healthcare, scientific and biomedical research, and education in the future.




\bibliographystyle{IEEEtran}

\vfill

\end{document}